\documentclass[twocolumn,twocolappendix]{aastex701}

\usepackage[T1]{fontenc}
\usepackage{lmodern}

\usepackage{amsmath}
\usepackage{graphicx}

\begin{document}

\title{
A Hycean Interpretation of K2-18b
Supported by Photochemical Atmospheric Compositional Structure
}

\correspondingauthor{Takuya Fujisawa}
\email{fujisawa@ep.sci.hokudai.ac.jp}

\author[orcid=0009-0000-6480-2668]{Takuya Fujisawa}
\affiliation{Department of Cosmosciences, Faculty of Science, Hokkaido University, Sapporo, Hokkaido 060-0810, Japan}
\email{fujisawa@ep.sci.hokudai.ac.jp}

\author[orcid=0009-0003-5023-7508]{Masashi Shimada}
\affiliation{Department of Cosmosciences, Faculty of Science, Hokkaido University, Sapporo, Hokkaido 060-0810, Japan}
\email{fujisawa@ep.sci.hokudai.ac.jp}

\author[orcid=0009-0003-0399-1305]{Tatsuya Yoshida}
\affiliation{Earth-Life Science Institute, Institute of Science Tokyo, 2-12-1 Ookayama, Meguro-ku, Tokyo 152-8550, Japan}
\email{tatsuya@tohoku.ac.jp}

\author[orcid=0000-0002-6757-8064]{Kiyoshi Kuramoto}
\affiliation{Department of Cosmosciences, Faculty of Science, Hokkaido University, Sapporo, Hokkaido 060-0810, Japan}
\email{keikei@ep.sci.hokudai.ac.jp}

\begin{abstract}
The nature of the sub-Neptune K2-18b is debated
between Hycean and mini-Neptune interpretations.
We test whether self-consistent Hycean atmospheres
are compatible with current JWST transmission spectra
by combining one-dimensional photochemical modelling,
radiative--convective equilibrium calculations, and
forward modelling of transmission spectra. We assume
H$_2$--CH$_4$--H$_2$O atmospheres over a liquid ocean,
compute altitude-dependent abundances with a 1D
photochemical model, and couple them to P--T profiles
that avoid runaway greenhouse states. Using the
CH$_4$-dominated 2.8--4.0\,$\mu$m band, we constrain
wavelength-independent offsets between NIRISS SOSS
and NIRSpec G395H for multiple reductions, and then
scan grids of CO and CO$_2$ scaling factors, weighted
by the CH$_4$-band offset posteriors, to evaluate
oxidised-carbon abundances consistent with the 4--5\,$\mu$m
region. Radiative--convective calculations further map
pressures and albedos that yield non–runaway climates.
Over a
wide range of temperatures and pressures, liquid oceans
can exist, and Hycean models with a 1\,bar H$_2$
envelope, percent-level CH$_4$ and CO, and CO$_2$
buffered at $\sim 10^{-3}$--$10^{-2}$ reproduce the NIRISS
and NIRSpec spectra from 0.8 to 5.2\,$\mu$m without
invoking DMS or other additional species. Our photochemical simulations show that H$_2$--CH$_4$--H$_2$O
networks generically drive CO to mixing ratios of
order 1--2\%. Mass-balance arguments imply that a
$\sim$1\,bar H$_2$ envelope with percent-level CH$_4$
requires interior replenishment on gigayear timescales,
and the resulting vertical gradients naturally generate
flat, CH$_4$-dominated plateaux in transmission. While
mini-Neptune scenarios remain viable, our results show
that Hycean configurations are likewise consistent with
the data, and current CO and CO$_2$ constraints alone
are not yet sufficient to rule out Hycean interpretations
of K2-18b.
\end{abstract}

\keywords{\uat{Exoplanet atmospheres}{487} --- \uat{Habitable planets}{695} --- \uat{Ocean planets}{1151} --- \uat{Transmission spectroscopy}{2133} --- \uat{Mini-Neptunes}{1063} --- \uat{Astrochemistry}{75} --- \uat{Biosignatures}{2018}}

\section{Introduction}\label{sec:Intro}

The search for habitable exoplanets has entered a new
era with the advent of the James Webb Space Telescope (JWST).
Sub-Neptunes, with radii between that of Earth and Neptune,
represent the most common type of exoplanets in the Milky Way,
yet they lack a Solar System analog \citep[e.g.,][]{Fulton2017, Rogers2015}. Whether these planets can
host life depends largely on their atmospheric and bulk properties.
The ``Hycean World'' (Hydrogen + Ocean) scenario has
been proposed as a habitable candidate, characterized
by a liquid water ocean beneath a hydrogen-rich atmosphere \citep{Madhusudhan_2021}.
Theoretical studies suggest that such worlds could maintain
liquid water over a wider range of orbital distances compared
to Earth-like planets \citep{Pierrehumbert2011, Madhusudhan2021}.

Among the detected exoplanets, K2-18b, a sub-Neptune orbiting an M2.5V dwarf star within the habitable zone, has drawn significant attention. Early observations by the Hubble Space Telescope implied the presence of water vapor \citep{Benneke2019, Tsiaras2019}. More recently, JWST transmission spectroscopy suggested the presence of carbon-bearing molecules, including CH$_4$ and CO$_2$, and a scarcity of NH$_3$ \citep{Madhusudhan2023}. These features are interpreted as potential evidence for a surface ocean beneath a hydrogen-rich atmosphere. Specifically, the observed depletion of NH$_3$ is theoretically attributed to its high solubility in liquid water, suggesting that the nitrogen reservoir resides in a surface ocean rather than the atmosphere \citep{Madhusudhan2021}. Furthermore, the capacity of an ocean to dissolve and buffer CO$_2$ is crucial for regulating the atmospheric CO$_2$ budget over long timescales, linking the atmospheric composition directly to the underlying ocean. The tentative detection of dimethyl sulfide (DMS), a molecule primarily produced by life on Earth, has further fueled the debate regarding the potential biosignatures on K2-18b \citep{Madhusudhan2023}. On Earth, DMS is known to play a crucial role in the climate system by acting as a precursor for cloud condensation nuclei (CCN), potentially increasing cloud albedo and regulating the surface temperature (the CLAW hypothesis; \citep{Charlson1987}).

However, recent independent reanalyses of the JWST
transmission spectra have presented a different perspective.
\citet{Schmidt2025} utilized multiple data reduction pipelines
and retrieval codes to reassess the atmospheric composition of K2-18b.
Their comprehensive study confirmed the robust detection of CH$_4$ but found
no statistically significant evidence for CO$_2$ or DMS, contrary to the
previous reports. Consequently, they argue that the observed spectrum
is better explained by a gas-rich ``mini-Neptune'' with an oxygen-poor
atmosphere, rather than a Hycean world. These contrasting interpretations underscore that the current observations do not yet uniquely constrain the presence or absence of CO$_2$. 

From theoretical view points, the Hycean interpretation inferred from the current transit spectra of K2-18b raises several physical and chemical issues, because K2-18b orbits an M-dwarf star and is exposed to strong ultraviolet irradiation. In particular, as discussed below, there are concerns about photochemical instability of the atmosphere, the long-term survival of a thin H$_2$ envelope, and the possibility of a runaway greenhouse state driven by the greenhouse effect of an H$_2$-rich atmosphere.

First, under Hycean conditions, previous studies have shown that an H$_2$-dominated atmosphere can maintain a warm surface in contact with a liquid ocean, with representative surface temperatures around 328~K \citep[e.g.,][]{Hu2021,Wogan2024}. For K2-18b, such a scenario corresponds to an H$_2$-dominated atmosphere containing CH$_4$ with H$_2$O simultaneously present. In this situation, radicals produced by H$_2$O photolysis oxidise CH$_4$ into CO and CO$_2$. This behaviour is difficult to reconcile with the fact that the transit spectra currently favoured by retrieval analyses do not show prominent spectral features of oxidised carbon species. In addition, under strong UV and X-ray irradiation, CH\textsubscript{4} itself is efficiently photodissociated.
Without recycling or a compensating source, CH\textsubscript{4} can therefore be destroyed on timescales much shorter than the age of the system.
Previous one-dimensional photochemical models have shown that maintaining the CH\textsubscript{4} levels inferred from observations may require substantial surface or interior fluxes, even if these fluxes are not biological in origin \citep{Wogan2024}.

Second, for a sub-Neptune like K2-18b orbiting close to its M-dwarf host star, it is not obvious that a thin H$_2$ envelope can survive over long timescales. In Hycean scenarios, the surface pressure of H$_2$ is typically assumed to be at most a few bar, and such a thin H$_2$ layer on a planet in a close-in orbit is expected to be vulnerable to hydrodynamic escape driven by high-energy stellar radiation. Even when we use the hydrodynamic escape formula H2 escape limited by CH4 cooling presented by \citet{Yoshida2024}, a 1~bar H$_2$ atmosphere would be lost in $\sim 4 \times 10^7$~yr in the absence of replenishment.

Third, the temperature structure of an H$_2$-rich atmosphere must be such that liquid water can exist at the bottom of the envelope without allowing the climate to enter a runaway greenhouse state. Collision-induced absorption by H$_2$–H$_2$ and H$_2$O–H$_2$O pairs gives H$_2$-dominated atmospheres a strong greenhouse effect, and it has been pointed out that, unless the planetary Bond albedo is sufficiently high due to enhanced reflection, the climate can become extremely hot or even undergo a runaway greenhouse transition \citep[e.g.,][]{Pierrehumbert2011,Innes2023}. 


Previous theoretical and observational studies have
addressed parts of these issues, but they have not yet
placed comprehensive constraints on Hycean scenarios
for K2-18b. In particular, under Hycean-like conditions
on K2-18b, CH$_4$ is expected to be oxidised by H$_2$O, lead-
ing to a continuous increase in CO$_2$, but it has not been
quantitatively assessed whether such CO$_2$ production
may in fact be buffered by the ocean. CO is likewise
expected to accumulate, yet no prominent CO features
have been detected to date---including in the analyses by
\citet{Madhusudhan2023} and \citet{Schmidt2025}---and neither these
nor any other spectral analyses have ruled out its presence. Furthermore, whether a very thin H$_2$ envelope can survive over long
timescales has so far been discussed only in terms of order-of-magnitude
escape rates, based on estimate such as the one above,
and the extent to which H$_2$ produced by photochemistry can modify the
atmospheric lifetime has not been explicitly evaluated for K2-18b. In addition, the range of H$_2$ surface
pressures and planetary albedos that yields warm but
non–runaway climates has not been examined systematically.
In particular, the surface temperature determined by
these conditions sets the depth and physical state of the
ocean beneath the envelope and thus controls the ability
of the ocean to dissolve and buffer CO$_2$. At the same
time, the tropopause pressure and temperature control the efficiency
of the cold trap and the H$_2$O mixing ratio in the upper
atmosphere, which in turn governs the production of ox-
idising radicals such as OH and affects the destruction
rate of CH$_4$ and the production rate of CO$_2$ and CO.

Previous Bayesian retrieval studies have adopted
a free-chemistry framework in which the abundances of
the main molecular species, the pressure–temperature
structure, cloud and haze parameters, and inter-instrument
offsets are all treated as free parameters
\citep[e.g.,][]{Madhusudhan2023,Schmidt2025}, without
explicitly incorporating the vertical gradients and chemical
couplings predicted for Hycean-like atmospheres under
strong M-dwarf irradiation. In particular, offsets between different instruments or
reductions have often been inferred in retrievals that
include molecules whose detections are themselves not
firmly established, raising the question of whether the
adopted offsets are always optimally chosen for each
reduced data set.

In this study, we investigate these unresolved problems
by combining one-dimensional photochemical modelling,
radiative–convective equilibrium calculations, and transit
spectrum analysis. First, using trial chemical abundances
motivated by the spectral features reported by
\citet{Madhusudhan2023}, we perform one-dimensional
photochemical simulations under the irradiation from an
M2.5V star to characterise the reaction network operating
in Hycean-type atmospheres. We then examine the production and accumulation of CO$_2$ and CO in an H$_2$–CH$_4$–H$_2$O atmosphere
as a function of the stratospheric H$_2$O abundance, in a series of lower-boundary-CO2-free experiments where no CO2 mixing ratio is imposed at the lower boundary. From these simulations, we diagnose the net
photochemical production flux of CO$_2$ and combine it with
a simple ocean-buffer model to estimate the range of
atmospheric CO$_2$ abundances that are compatible with
long-term CO$_2$ dissolution into a deep ocean.

We will also re-examine the lifetime of CH\textsubscript{4} within our photochemical framework.
At the same time, by combining hydrodynamic escape rates taken from the literature with the H\textsubscript{2} production rates obtained in this study, we estimate the residence time of a thin H\textsubscript{2} envelope.

Next, we use the photochemically predicted vertical
profiles to compute transmission spectra. In doing so, we
fit the CH$_4$-dominated 2.8–4.0\,$\mu$m band for each
reduced data set to infer the range of wavelength-independent
offsets between NIRISS and NIRSpec that remains
consistent with the robustly detected CH$_4$ features.
We then scan grids of CO and CO$_2$ scaling factors,
weighting each model by the posterior distribution of
the offset derived from the CH$_4$ band, in order to
evaluate which combinations of oxidised carbon abundances
are compatible with the 4–5\,$\mu$m region.

Finally, we employ radiative–convective equilibrium
calculations to determine the temperature structure of
H$_2$-rich atmospheres over liquid H$_2$O oceans and map
the combinations of H$_2$ surface pressure and planetary
Bond albedo that yield warm but non–runaway climates.
By relating the resulting tropopause temperatures and
stratospheric H$_2$O mixing ratios to the parameter space
explored in our photochemical model, we clarify how
Hycean climate states connect to the CO$_2$ mass balance.
Taken together, these steps allow us to assess to what
extent Hycean compositions that are consistent with
photochemistry and climate can also reproduce the
current JWST transmission spectra of K2-18b.

\section{MODEL DESCRIPTION}

\subsection{Photochemical Model}
\label{sec: PM}

We apply PROTEUS, a one-dimensional photochemical model for a plane-parallel atmosphere \citep{Nakamura2023a}. PROTEUS has previously been applied to the Jovian ionosphere \citep{Nakamura2022}, the present-day Martian atmosphere \citep{Nakamura2023a,Yoshida2023}, an early Martian atmosphere, an H$_2$O-dominated atmosphere in the runaway-greenhouse regime \citep{Kawamura2024}, and the photochemical evolution of the early Earth’s atmosphere \citep{Yoshida2024}. The details of PROTEUS are described in \citet{Nakamura2023a}; below, we summarize the setup used in this study.

For the chemical processes, we consider a neutral H–C–O network composed of 48 species. The reactive species are H$_2$O, O($^1$D), OH, H$_2$, H, O$_3$, O$_2$, HO$_2$, O, H$_2$O$_2$, CO$_2$, CO, C, HCO, H$_2$CO, CH$_4$, CH$_2$, $^1$CH$_2$, CH$_3$, C$_2$H$_5$, C$_2$H$_2$, C$_2$H, C$_2$, C$_2$H$_4$, C$_3$H$_8$, C$_3$H$_7$, C$_2$H$_3$, C$_3$H$_6$, CH, CH$_2$CO, CH$_3$CHO, C$_2$H$_5$CHO, C$_3$H$_3$, C$_3$H$_2$, CH$_3$C$_2$H, CH$_2$CCH$_2$, C$_3$H$_5$, CH$_3$O$_2$, CH$_3$CO, C$_2$H$_2$OH, C$_2$H$_4$OH, CH$_3$O, CH$_2$, CH$_2$OH, CH$_3$OH, H$_2$CCO, and HCCO. In total, 322 reactions (bimolecular, termolecular, and photolysis) are included. 

Reactions involving N- and S-bearing species are not considered in this study, because, following the Hycean scenarios discussed by \citet{Hu2021}, highly soluble species such as NH$_3$ and H$_2$S are expected to partition efficiently into a surface ocean. We therefore assume that most of the nitrogen and sulfur reservoirs reside in the underlying ocean rather than in the atmosphere, so that the impact of N and S chemistry. Molecular nitrogen (N$_2$) is included as a background gas with a surface mixing ratio of 1\% and is treated as chemically inert in this study; it does not change chemically but participates in the reaction network as a third body in termolecular reactions. 

Our reaction network is based primarily on the H–C–O schemes of \citet{Tian2011} and \citet{Shang2017}, augmented with additional reactions required to describe C$_2$ and C$_3$ hydrocarbon chemistry up to C$_2$H$_6$ and C$_3$H$_8$, which are produced photochemically from methane. We explicitly include C$_2$H$_6$ because it can contribute to near-infrared opacity in our transmission spectrum calculations, and C$_3$H$_8$ as a representative precursor of higher-order organics that participate in the formation of an organic haze layer. To improve the accuracy of carbon chemistry under UV irradiation, we incorporate updated photodissociation cross-sections for CO, utilizing the absolute optical oscillator strengths and photoabsorption data (7–200 eV) provided by \citet{Chan1993}. The complete list of reactions, including rate coefficients, is provided as online supplementary material hosted at Zenodo \citep{OurZenodo2026}.

\begin{figure}
 \centering
 \includegraphics[width=\columnwidth]{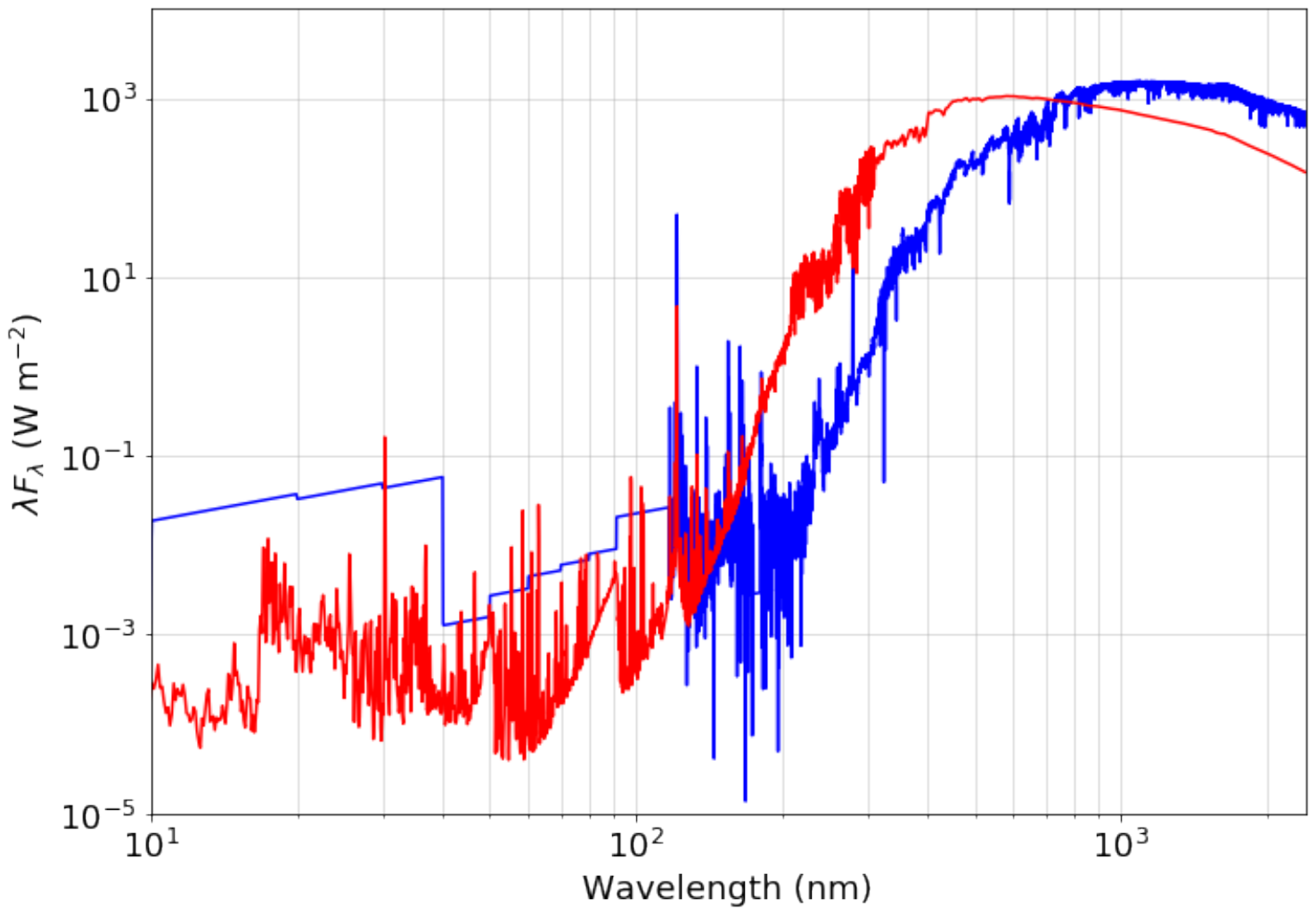}
 \caption{Assumed spectrum based on the observation for a M2.5V dwarf star GJ176 (blue line). The intensity at each wavelength is converted to the value correspond to the orbit of K2-18b. For comparison, the solar spectrum is shown by the red line, scaled to match the total bolometric flux. The vertical axis represents $\lambda F_\lambda$, meaning that the area under the curve in this logarithmic plot is proportional to the integrated energy flux.}
\label{fig:stellar_spectrum}
\end{figure}

Stellar spectrum of K2-18 is not available in the UV range critical for photochemistry, so we adopt the observed spectrum of the M2.5V dwarf GJ~176 (Figure~\ref{fig:stellar_spectrum}). The spectrum is scaled to the orbital distance of K2-18b. Since K2-18 and GJ~176 are both classified as M2.5V dwarfs and exhibit comparable levels of stellar activity, this substitution provides a reasonable representation of the UV–visible radiation environment relevant for photochemistry \citep[e.g.,][]{Hu2021,Wogan2024}.

\begin{figure}
 \centering
 \includegraphics[width=\columnwidth]{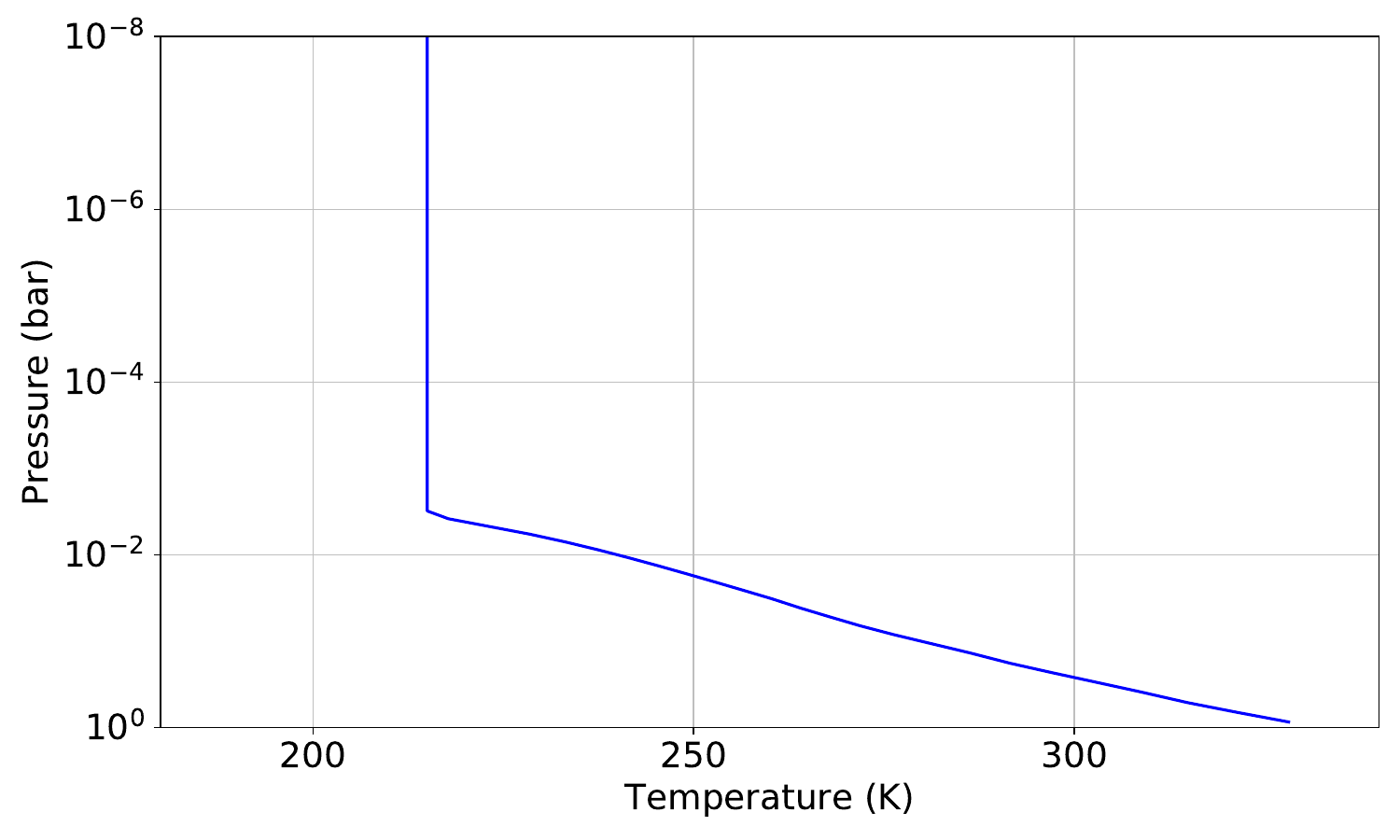}
 \caption{Pressure–temperature (P–T) profiles for the K2-18b
    atmosphere. This panel shows the Hycean
    K2-18b P–T profile adopted in this work: the surface
    temperature is 328~K and the surface pressure is
    1~bar. The troposphere follows an adiabatic temperature
    gradient up to the tropopause, above which the
    stratosphere is isothermal at 215~K.  
}
\label{fig:pt_profile}
\end{figure}

We consider an H$_2$-dominated atmosphere representative of Hycean interpretations for K2-18b. Throughout the photochemical calculations, the H$_2$ abundance is kept fixed (i.e., we do not explicitly include upper-atmospheric escape of H$_2$). We represent the thermal structure with a 1-bar H$_2$-rich envelope above a liquid-water ocean and consider a surface temperatures $T_{\rm s}=328$ K (Figure \ref{fig:pt_profile}). This choice is made to ensure consistency with previous Hycean modeling studies for temperate sub-Neptunes \citep{Hu2021,Wogan2024}. 

In the photochemical calculations we focus on atmospheres with a surface pressure of 1 bar. This value is broadly consistent with temperate Hycean-like solutions found in our radiative–convective framework, in which substantially higher H$_2$ surface pressures at the same Bond albedo tend to push the climate toward much warmer states where temperate surface conditions are no longer realised (Section \ref{sec:climate_stability}). At the same time, the key photochemical processes that control the mass balance of H, C, and O in our model operate primarily in the low-pressure stratosphere, so that the steady-state abundances and mass fluxes of the major species are only weakly sensitive to moderate changes in surface pressure within the Hycean regime. 

The troposphere follows an adiabatic temperature profile for an H$_2$–H$_2$O mixture up to the tropopause, above which the stratosphere is assumed to be isothermal at 215 K. This stratospheric temperature is close to the skin temperature implied by the outgoing longwave radiation in our radiative–convective equilibrium solutions (Section \ref{sec:climate_stability}) and falls within the range of equilibrium stratospheric temperatures obtained for similar H$_2$-rich temperate sub-Neptunes \citep{Hu2021}. The atmosphere is divided into 100 layers, equally spaced in altitude from the surface to 900 km, corresponding to pressures down to about $10^{-8}$ bar (Figure \ref{fig:pt_profile}).

For vertical transport, we assume a constant eddy diffusion coefficient of $5\times10^{5}\ {\rm cm^2\ s^{-1}}$ throughout the atmosphere, following the simple parameterization adopted by \citet{Wogan2024} for temperate sub-Neptunes.  

H$_2$O is assumed to condense to form a liquid ocean at the surface. In the troposphere, the H$_2$O vapour mixing ratio is prescribed to follow the saturation curve up to the tropopause. Above the tropopause, H$_2$O is treated as a time-dependent species and is initialised with a uniform stratospheric mixing ratio of $10^{-6}$, a value comparable to typical Earth stratospheric H$_2$O abundances and indicative of an atmosphere in which the cold trap efficiently limits the upward transport of water.

At the lower boundary, we additionally allow for deposition of soluble formaldehyde (H$_2$CO) into the ocean, applying a surface deposition velocity of $1.0\times10^{-3}\ \mathrm{m\,s^{-1}}$, consistent with recent applications of PROTEUS (e.g., \citealt{Koyama2024a}). This boundary condition links the vertical transport of H$_2$CO to efficient scavenging by the ocean surface and prevents its artificial accumulation in the lowest atmospheric layers.

To represent the present atmospheric state inferred from the latest JWST retrievals by \citet{Madhusudhan2023}, we first explore a CO$_2$-bearing configuration in which the surface mixing ratios of both CH$_4$ and CO$_2$ are fixed at 1\%. In this framework, the fixed 1\% surface mixing ratio of CH$_4$ represents the currently observed abundance and allows us to calculate steady-state vertical profiles of all derived species consistent with that CH$_4$ level; the long-term stability of CH$_4$---specifically, whether surface or interior sources can balance the computed photochemical loss rate---is then assessed a posteriori in the mass-balance analysis (Section~\ref{sec:photochemi} and \ref{sec:CH4,H2}). We refer to this setup as the
\emph{CO$_2$-bearing reference case} throughout this paper. 

The fixed 1\% surface mixing ratio of CO$_2$ is assumed to represent dissolution equilibrium with an underlying ocean--interior reservoir that maintains a nearly constant surface partial pressure of CO$_2$ over the timescales of interest, rather than a prescribed constant upward flux. At the start of the integration, CH$_4$ and CO$_2$ are uniformly distributed with altitude at their surface mixing ratios, and their vertical profiles subsequently evolve under the combined action of chemistry and vertical transport, subject to these fixed lower boundary conditions. In this setup with fixed lower-boundary mixing ratios of CH$_4$ and CO$_2$, we regard the point at which the abundances of all chemical species cease to evolve with time as the steady state.

In addition to this CO2-bearing setup, we also perform a set of experiments in which the CO2 mixing ratio is set to zero throughout the atmosphere at the start of the integration and no CO2 lower-boundary flux is imposed, following the CO2-poor scenarios suggested by \citet{Schmidt2025}. These runs are designed as lower-boundary-CO2-free experiments, motivated by the possibility that atmospheric CO2 can be buffered by dissolution into a deep ocean. In these calculations, CH4 is again fixed at the surface, but the H2O mixing ratio in the stratosphere above the tropopause is treated as a free parameter and prescribed to be vertically uniform in the range 10$^{-7}$–10$^{-3}$. For the water vapor abundance in the troposphere, we assume, for simplicity, that it varies continuously from the tropopause
value; therefore, the pressure and temperature at the tropopause are not
varied. Throughout this work, the stratospheric H$_2$O mixing ratio used
in the photochemical and mass-balance calculations is defined as the
value at the cold trap near the tropopause. 

In the lower-boundary-CO2-free experiments, atmospheric CO$_2$ is produced photochemically and, in this system, its abundance would continue to increase with time so that no global steady state is obtained for CO$_2$. In the time-dependent calculations, we therefore identify the point at which the abundances of all species other than CO$_2$ no longer change with time and define this condition as the equilibrium state. The long-term accumulation level of CO$_2$ in the atmosphere inferred from these experiments is discussed in detail in Section~\ref{sec:CO2_H2O_dependence} and Appendix~\ref{sec:appendix_CO2mass}. These runs are intended as idealised experiments without an imposed CO$_2$ source at depth, rather than as a claim that K2-18b is necessarily CO$_2$-free.

For both the CO$_2$-bearing reference case and these lower-boundary-CO2-free experiments, we adopt the same fiducial Hycean temperature–pressure profile shown in Figure~\ref{fig:pt_profile}, in order to investigate the dependence of the CO$_2$ mass flux on the prescribed stratospheric H$_2$O abundance alone, disentangled from changes in surface temperature or total H$_2$ pressure.

\subsubsection{Global Mass-Balance Diagnostics} \label{sec:CH4_mass_balance}

In order to evaluate the long-term stability of the atmospheric composition, we perform global mass-balance diagnostics in two steps.
For each experiment, we first compute the steady-state atmospheric composition defined by the adopted lower-boundary conditions.
We then conceptually relax these lower-boundary constraints and diagnose the globally integrated production and loss rates that would result, using these rates to infer characteristic depletion timescales.
The unit of these fluxes is kg s$^{-1}$, and they are defined by integrating the local production and loss rates over the entire volume of the plane-parallel atmosphere.
From these fluxes, we compute the total molar balance of H, C, and O and verify that mass conservation is satisfied to numerical accuracy.

\subsection{Transit Spectrum Modeling}
\subsubsection{Transit depth calculation method}
\label{sec:transit_method}
\begin{figure}
 \centering
 \includegraphics[width=\columnwidth]{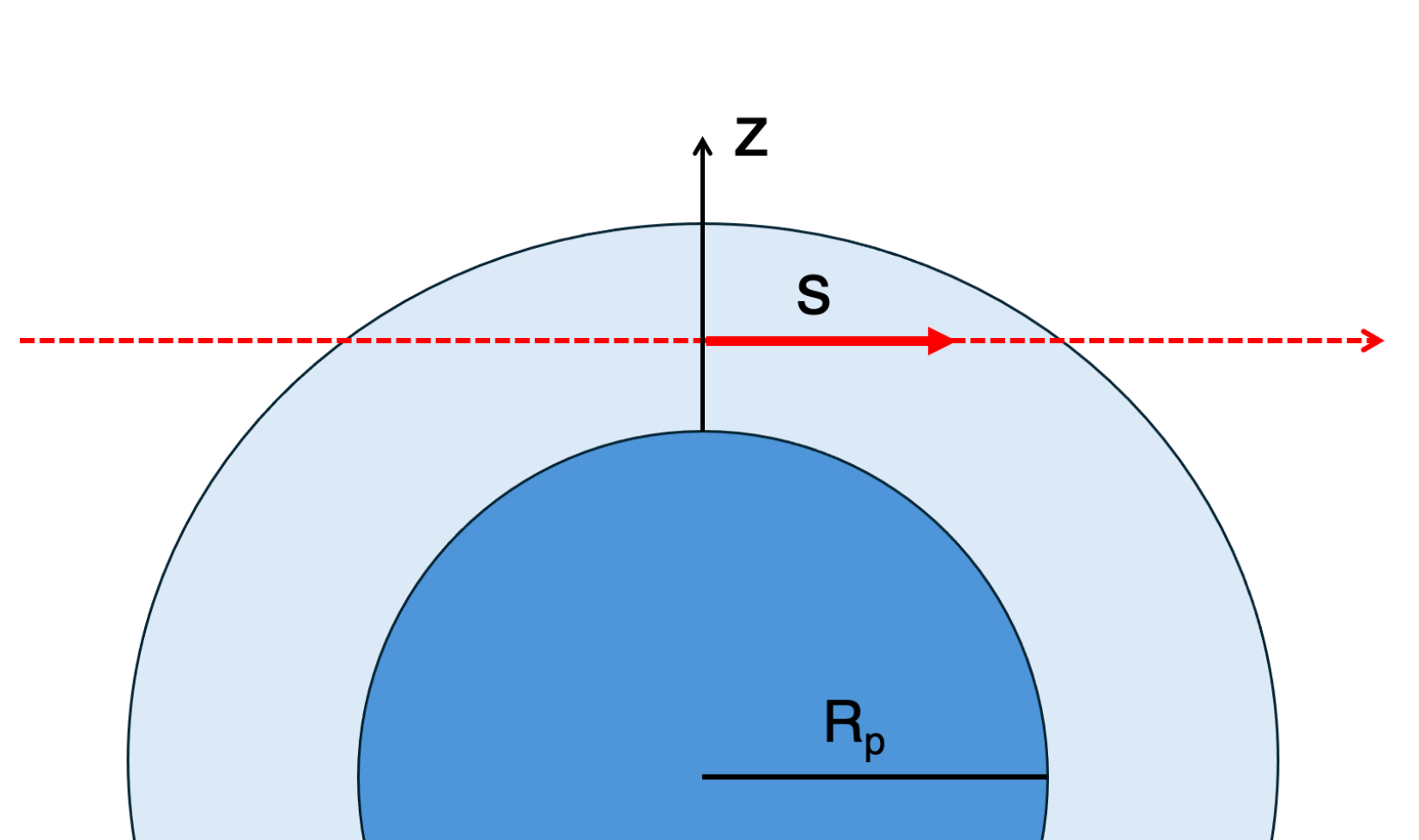}
 \caption{Conceptual illustration of a light ray passing through the planetary atmosphere at altitude $z$ from the planetary surface. The blue region represents the planet, and the light blue region represents the atmosphere. The red arrow indicates a light path passing through the atmosphere. The altitude of the light path from the planetary surface is denoted by $z$, and coordinates $s$ are taken along the path. The origin of the $s$-coordinate is set at the point of the lowest altitude on the light path.}
\label{fig:transit_geometry}
\end{figure}

Along the light path passing through altitude $z$ from the planetary surface (Figure~\ref{fig:transit_geometry}), we define the coordinate $s$ along the path. The optical depth $\tau_{\lambda}$ across the total light path at wavelength $\lambda$ is expressed as
\begin{align}\label{eq:optical_depth}
\tau_{\lambda}(z) &= \kappa_i(z)\,n_i(z)\,
\exp\left(-\frac{z}{H}\right)
\sqrt{\frac{2(R_p+z)H}{\pi}},
\end{align}%
where $n_i(z)$ is the number density of each chemical species and $\kappa_{\lambda,i}(z)$ is the absorption coefficient of each species at wavelength $\lambda$. $\tau_\lambda(z)$ is the slant optical depth along the path at impact parameter $z$, $R_p$ is the planetary radius at the reference surface, and $R_s$ is the stellar radius.

 The apparent radius of the planet, $R_{\lambda, \text{trans}}$, at wavelength $\lambda$ is expressed as follows:
\begin{align}
S_{\text{atm}} &= \sum_{i=0}^{100}\pi[(R_p+z_{i+1})^2-(R_p+z_i)^2] \label{eq:S_atm}, \\
S_{\lambda, \text{trans}} &= \sum_{i=0}^{100}\pi[(R_p+z_{i+1})^2-(R_p+z_i)^2]\exp[-\tau_{\lambda}(z_i)] \label{eq:S_trans}, \\
R_{\lambda, \text{trans}} &= \sqrt{R_p^2+\frac{(S_{\text{atm}}-S_{\lambda, \text{trans}})}{\pi}} \label{eq:R_trans_calc}.
\end{align}%
where $S_{\text{atm}}$ is the total cross section of the region where the atmosphere is present, and $S_{\lambda,\text{trans}}$ is the total cross section transparent for atmospheric absorption.
In Equation~(\ref{eq:S_trans}),
$\exp[-\tau_\lambda(z_i)]$ represents the fraction of stellar light
transmitted through the atmospheric annulus at impact parameter $z_i$.

Using the apparent radius $R_{\lambda, \text{trans}}$, the transit depth is expressed as follows:
\begin{equation} \label{eq:transit_depth_final}
\frac{R_{\lambda, \text{trans}}^2}{R_s^2}.
\end{equation}

\begin{figure*}
 \centering
 \includegraphics[width=\textwidth]{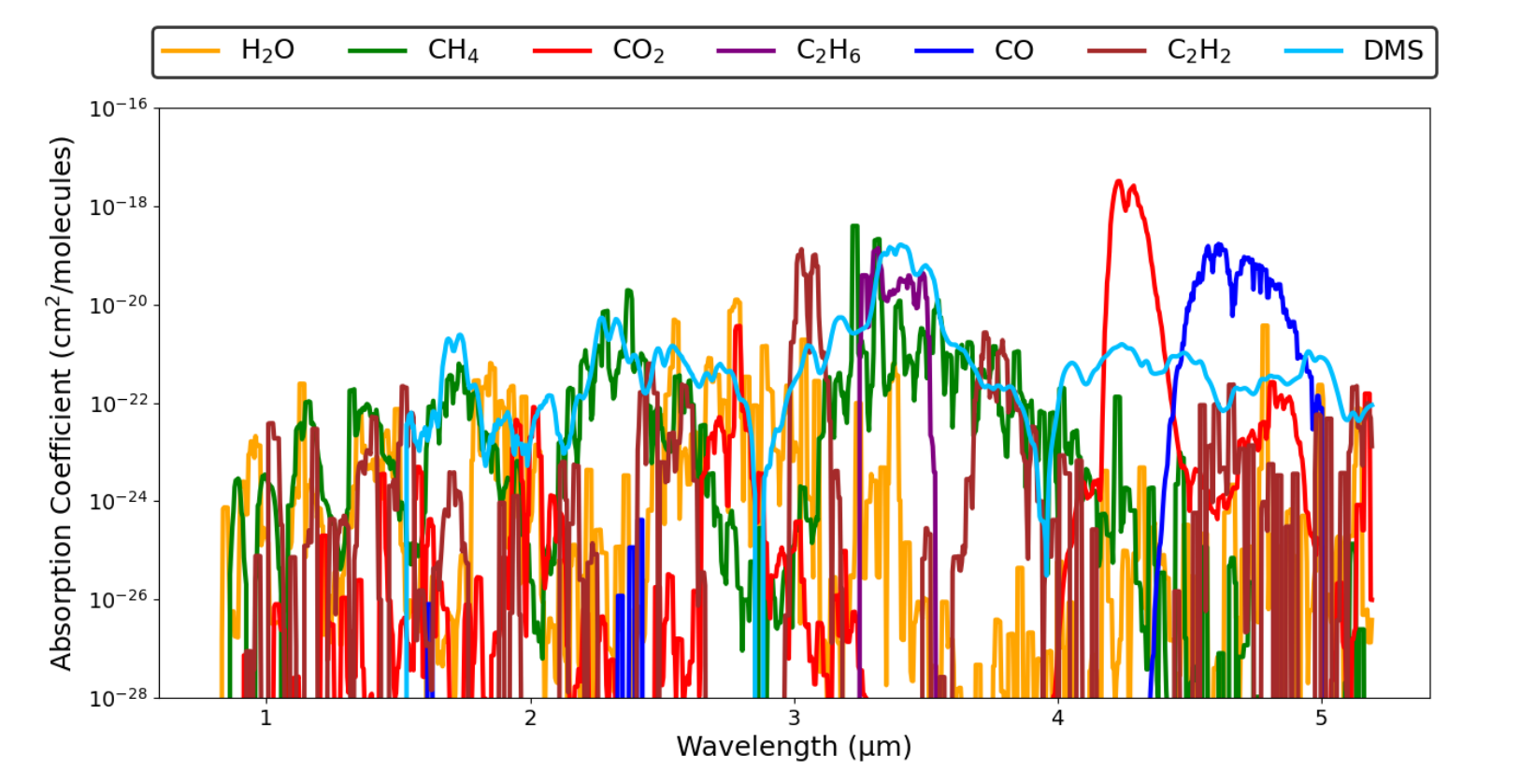}
 \caption{Graph of absorption coefficients used for calculating the transit depth. For improved clarity, a moving average has been applied to the absorption cross-sections. DMS data was obtained from HITRAN absorption cross section, considering N\textsubscript{2} atmosphere and a temperature-pressure broadening at 298 K and 1 bar. Information for other absorption coefficients was obtained from HITRAN HAPI, considering an H\textsubscript{2} background atmosphere and the P-T profile for each altitude.}
\label{fig:opacity_spectra}
\end{figure*}

Absorption coefficients for H$_2$, CO, CO$_2$, CH$_4$, C$_2$H$_2$, and C$_2$H$_6$ were obtained from HITRAN HAPI \citep{Kochanov2016}. For DMS, we used absorption cross-sections from the HITRAN database computed for an Earth-like atmosphere (298~K, 1~bar). The wavelength-dependent absorption cross-sections used in the transit spectrum calculations are
summarised in Figure~\ref{fig:opacity_spectra}.

Considering pressure broadening by the H$_2$-dominated background atmosphere, the temperature and pressure at each altitude were taken from the P–T profile shown in Figure~\ref{fig:pt_profile}. In the forward modeling, this atmospheric temperature–pressure structure was kept fixed rather than treated as a free parameter, in order to (i) enforce consistency with the Hycean configuration assumed in our photochemical and radiative–convective models, and (ii) avoid additional degeneracies between temperature, composition, and cloud properties when interpreting the transmission spectra. This choice allows us to isolate the impact of the vertical mixing-ratio structure on the spectral shape. The validity of adopting this fixed P–T profile is supported by our radiative–convective equilibrium calculations presented in Section~\ref{sec:climate_stability}, which show that, for the same H$_2$-rich compositions considered here, self-consistent radiative–convective solutions converge to temperature structures closely resembling the profile in Figure~\ref{fig:pt_profile} over the pressure range probed by transmission spectroscopy.

The atmospheric temperature–pressure profile was therefore fixed to that of Figure~\ref{fig:pt_profile} in all transit spectrum calculations. The reference planetary radius $R_p$ in our model is defined at the nominal “surface” (ocean top), and the wavelength-dependent effective transit radius $R_{\lambda,\mathrm{trans}}$ is obtained from the atmospheric cross sections as described in Section~\ref{sec:transit_method}. The model transit depth is then given by $R_{\lambda,\mathrm{trans}}^2 / R_s^2$, where $R_s$ is the stellar radius. We use the planet-to-star radius ratio $(R_p/R_s)$ reported by \citet{Madhusudhan2023} as the baseline for the absolute transit depth, and allow only a small adjustment of $R_p$ within the observational uncertainty to align the overall model level with the data. This adjustment does not affect the gravity used in our photochemical or radiative–convective models and therefore does not influence the atmospheric structure or chemistry.

For other atmospheric absorption and scattering processes, we adopted the parameterization from the AURA transit code \citep{Pinhas2018}.
Specifically, our retrieval model includes parameters for Rayleigh scattering (enhanced by a factor $a$ and with a slope $\gamma$), a grey cloud deck characterized by a cloud-top pressure ($P_c$), and a fractional cloud/haze coverage ($\phi$).
These parameters, along with their prior distributions listed in Table \ref{tab3}, allow us to account for the impact of aerosols and atmospheric opacity sources beyond the molecular absorption .
Stellar contamination was approximated by a blackbody function, and we considered absorption suppression by clouds, absorption by haze, and collision-induced absorption CIA by H\textsubscript{2}-H\textsubscript{2} collisions. These parameters are listed in Table \ref{tab1}.

\begin{deluxetable*}{lcccccccc}
\tablecaption{Fitted temperature and cloud/haze properties and stellar
heterogeneity parameters for K2-18b, for the four data reductions used
in this work. All cases adopt the same Hycean temperature--pressure
profile, and the vertical mixing ratios follow \citet{Madhusudhan2023}
(uniform) or the photochemical profiles from this study, as described in
Section~2.2.\label{tab1} These best-fit
values are obtained from our full forward-model fits to the combined
NIRISS+NIRSpec data and are held fixed for each reduction in the subsequent
CO/CO$_2$ grid scan described in Section~\ref{subsubsec:method_CO}.}
\tablehead{
\colhead{Reduction} &
\colhead{$\phi$} &
\colhead{$\log a$} &
\colhead{$\gamma$} &
\colhead{$\log(P_c/{\rm bar})$} &
\colhead{$T_{\rm phot}$ (K)} &
\colhead{$T_{\rm het}$ (K)} &
\colhead{$f_{\rm het}$} &
\colhead{$R_p/R_s$}
}
\startdata
Madhusudhan et al.~(2023) & 0.63 & 8.20 & $-11.11$ & $-0.51$ & 3600 & 3200 & 0.05 & 0.05229 \\
exoTEDRF                  & 0.63 & 8.20 & $-11.11$ & $-0.51$ & 3600 & 3200 & 0.05 & 0.05229 \\
FIREFLy                   & 0.63 & 8.20 & $-11.11$ & $-0.51$ & 3600 & 3200 & 0.05 & 0.05229 \\
Eureka! Reduction A                 & 0.63 & 8.20 & $-11.11$ & $-0.51$ & 3600 & 3200 & 0.05 & 0.05221 \\
Eureka! Reduction B                 & 0.63 & 8.20 & $-11.11$ & $-0.51$ & 3600 & 3200 & 0.05 & 0.05245 \\
\enddata
\end{deluxetable*}

\subsection{Transit-spectrum fitting framework}\label{subsec:method_transit}

\subsubsection{$\mathrm{CH_4}$-band offset scan}\label{subsubsec:method_offsets}

We use the CH$_4$-dominated wavelengths to delimit the
allowed absolute offsets between the model and each
reduced data set. For each reduction (the
\citet{Madhusudhan2023} reduction, \texttt{exoTEDRF} and \texttt{FIREFLy} from
\citealt{Schmidt2025}), we adopt the same Hycean,
photochemical profile as in Section~2.2.2 and restrict
the comparison to the $2.8$–$4.0~\micron$ range, where
CO and CO$_2$ contribute only weakly and CH$_4$ features
remain prominent.

For a given reduction, we define a wavelength-independent
offset in transit depth, $\Delta_{\rm off}$, such that the
model transit depth $d^{\rm mod}$ is compared to the
shifted data $d^{\rm obs}+\Delta_{\rm off}$. We then scan over a grid of \(\Delta_{\rm off}\) values from \(-100~\mathrm{ppm}\) to \(100~\mathrm{ppm}\) and compute
\begin{equation}
\chi^2_{\rm CH_4} =
\sum_{i}
\left[
\frac{d^{\rm obs}_i + \Delta_{\rm off} - d^{\rm mod}_i}
     {\sigma_i}
\right]^2,
\label{eq:chi2_CH4}
\end{equation}%
where the sum runs over all spectral bins in the
$2.8$–$4.0~\micron$ interval and $\sigma_i$ is the
observational uncertainty in bin $i$.

From $\chi^2_{\rm CH_4}$ we define the offset posterior as
\begin{equation}
p(\Delta_{\rm off}) \propto
\exp\left[-\frac{\Delta\chi^2_{\rm CH_4}}{2}\right],
\label{eq:post_offset}
\end{equation}%
where $\Delta\chi^2_{\rm CH_4} =
\chi^2_{\rm CH_4}-\chi^2_{{\rm CH_4},{\rm min}}$ and
$\chi^2_{{\rm CH_4},{\rm min}}$ is the minimum value over
the scanned range. We normalise $p(\Delta_{\rm off})$
such that $\int p(\Delta_{\rm off})\,{\rm d}\Delta_{\rm off}=1$
and adopt the central $68\%$ and $95\%$ credible
intervals as the $1\sigma$ and $2\sigma$ ranges of allowed
offsets.

In the following, we use the $R\approx 55$ binning for the
\citet{Madhusudhan2023} reduction and the $R\approx 100$
binnings for \texttt{exoTEDRF} and \texttt{FIREFLy} as
our baseline when quoting offset constraints. For the \texttt{Eureka!} reductions, no simultaneous NIRISS SOSS data are available. Instead, we vary the white-light baseline transit depth by adjusting the planet-to-star radius ratio within the uncertainties listed in Table~\ref{tab1}, and choose the value that best matches the CH$_4$-dominated band at 2.8–4.0\,$\mu$m. For this reason, when analysing the \texttt{Eureka!} data sets we keep $\Delta_{\mathrm{off}} = 0$ fixed.

\subsubsection{$\mathrm{CO}$ and $\mathrm{CO_2}$ retrievals}\label{subsubsec:method_CO}

Using the offset posteriors derived in
Section~\ref{subsubsec:method_offsets}, we next constrain
the CO abundance from the $4$–$5~\micron$ region,
where the CO fundamental band dominates. For each
reduction, we compute forward-model transmission
spectra using the Hycean temperature–pressure profile
and the photochemical reference composition, and we
scale the CO and CO$_2$ mixing ratios by multiplicative
factors defined on a fixed grid $(X_{\rm CO}, X_{\rm CO_2})$.
At each grid point, the model spectrum is binned to the
resolution of the $4.0$–$5.0~\micron$ data, and we
evaluate the contribution from this CO/CO$_2$-sensitive
region as
\begin{equation}
\chi^2_{\rm CO} =
\sum_{j}
\left[
\frac{d^{\rm obs}_j + \Delta_{\rm off} - d^{\rm mod}_j}
     {\sigma_j}
\right]^2,
\label{eq:chi2_CO}
\end{equation}%
where the sum runs over all spectral bins in the
$4.0$–$5.0~\micron$ interval and $d^{\rm mod}_j$ is the
model depth at bin $j$ for the chosen
$(X_{\rm CO}, X_{\rm CO_2})$. For the
\citet{Madhusudhan2023} reduction we use $R\approx 55$
binned data so that the number of spectral bins in the
CO band matches that of the other reductions, for which
we adopt $R\approx 100$.

The joint likelihood is then constructed by combining
the CH$_4$-band information and the CO band as
\begin{equation}
\chi^2_{\rm tot} =
\chi^2_{\rm CH_4} + \chi^2_{\rm CO},
\label{eq:chi2_tot}
\end{equation}%
and on the three-dimensional grid
$(\Delta_{\rm off}, X_{\rm CO}, X_{\rm CO_2})$ we assign
weights
\begin{equation}
w \propto
\exp\left[-\frac{\chi^2_{\rm tot}-\chi^2_{\rm min}}{2}\right],
\label{eq:weights}
\end{equation}%
where $\chi^2_{\rm min}$ is the global minimum over the
grid. We then marginalise over $\Delta_{\rm off}$ and
$X_{\rm CO_2}$ to obtain a one-dimensional posterior
for CO,

We adopt log-uniform priors for the CO and CO$_2$ mixing ratios over the ranges
$X_{\mathrm{CO}}, X_{\mathrm{CO_2}} \in [10^{-13},\,0.18]$. The quoted probabilities
$P(X_{\mathrm{CO}} \ge 10^{-3})$ and $P(X_{\mathrm{CO}} \ge 10^{-2})$ should therefore
be interpreted as posterior probabilities conditional on this prior support.

\begin{equation}
p(X_{\rm CO}) \propto
\sum_{\Delta_{\rm off}}\sum_{X_{\rm CO_2}} w,
\label{eq:post_CO}
\end{equation}%
which we normalise on the discrete logarithmic grid such that
$\sum_i p\bigl(X_{{\rm CO},i}\bigr) = 1$, where $X_{{\rm CO},i}$ are the grid
points in $X_{\rm CO}$. From
$p(X_{\rm CO})$ we derive, for each reduction, the
maximum-posterior (best-fit) value $X_{\rm CO,\,best}$,
the two-sided $95.4\%$ credible interval (corresponding
to $2\sigma$ for a Gaussian), and the integrated
probabilities $P(X_{\rm CO}\ge 10^{-3})$ and
$P(X_{\rm CO}\ge 10^{-2})$ used in
Section~\ref{subsec:CO_results}. An analogous procedure
is applied to construct posteriors for CO$_2$,
$p(X_{\rm CO_2})$, by marginalising over
$(\Delta_{\rm off}, X_{\rm CO})$.

In our CO/CO$_2$ grid scan we vary only the CO and CO$_2$ mixing ratios and
the wavelength–independent NIRSpec–NIRISS offset. All other atmospheric,
cloud/haze, and stellar-contamination parameters are fixed, for each
reduction, to the best-fit values listed in Table~1, which were obtained
from our full forward-model fits to the combined NIRISS+NIRSpec data using
the vertically structured Hycean atmosphere described in Section~2.2. We
therefore do not re-sample these parameters, nor assign explicit priors to
them, in the CO/CO$_2$ analysis presented in Sections~\ref{subsubsec:offset_results}–\ref{subsec:CO_results}.

\subsection{Radiative-Convective Model}\label{sec:rc-model}
We performed one-dimensional radiative–convective equilibrium calculations for an H$_2$-rich atmosphere on a liquid H$_2$O ocean, using the model developed by \citet{Yoshida2025}.
The model treats a plane-parallel atmosphere in hydrostatic balance, using the same gravity and mean molecular weight as in the photochemical model, and includes the major radiatively active species identified there (H$_2$, H$_2$O, CO$_2$, CH$_4$, CO, and C$_2$H$_6$). The atmosphere is discretized into 200 layers on a logarithmically spaced pressure grid from 1~bar down to $10^{-6}$~bar, and the corresponding altitude profile is obtained from hydrostatic balance.

We evaluate the integral of the radiative flux over a wavenumber range from 1 to 30,000 cm$^{-1}$, with a resolution of 1cm$^{-1}$. To optimize computational efficiency, gaseous absorption and Rayleigh
scattering are represented by pre-computed opacity tables. These tables were generated on a grid consisting
of 41 temperature points from 200 to 400 K in 5 K steps
and 10 logarithmically spaced pressure points from 10$^{-6}$
to 1.0 bar, and absorption cross-sections in the radiative–
convective calculations are obtained by interpolation on
this grid. Rayleigh scattering is included for H$_2$ and H$_2$O, with cross-sections taken from the implementation of the radiative–convective model by \citet{Yoshida2025}. The surface albedo is set to 0.06, assuming a dark ocean surface \citep{Hu2021}.

For gaseous absorption, we include collision-induced absorption (CIA) of H$_2$–H$_2$ and the H$_2$O continuum absorption, as well as line absorption by H$_2$O, CH$_4$, CO$_2$, CO, and C$_2$H$_6$. H$_2$O line data are taken from the HITRAN2020 database \citep{Gordon2022} using the HITRAN Application Programming Interface \citep[HAPI;][]{Kochanov2016}. The H$_2$O line profile is assumed to follow a Voigt profile, truncated at 25~cm$^{-1}$ from the line center, and combined with continuum absorption represented by the MT\_CKD~4.3 model \citep{Mlawer2023} without double counting. This setup allows us to capture the greenhouse contributions of the major photochemical species predicted in our 1D chemistry calculations.

Convection is represented by imposing an adiabatic temperature profile in the troposphere. At each iteration, the lapse rate is computed from the mixture heat capacity $C_p(T)$, including its temperature dependence, for the H$_2$–H$_2$O–CO$_2$–CH$_4$–CO–C$_2$H$_6$ mixture. Where H$_2$O is close to saturation, the effect of latent heat release is included to obtain a moist-adiabatic lapse rate. Heat capacity data are taken from the NIST-JANAF Thermochemical Tables \citep{Chase1998} via the NIST Chemistry WebBook (NIST SRD 69) and represented by Shomate polynomials. 

The upward and downward radiative fluxes are computed by solving the
two-stream radiative transfer equations used in \citet{Yoshida2025},
with the gaseous absorption and Rayleigh scattering described above. For a given trial P–T profile, radiative–convective equilibrium is sought by iteratively adjusting the stratospheric temperature and the vertical temperature structure such that the top-of-atmosphere outgoing longwave radiation balances the globally averaged absorbed stellar flux. The troposphere is assumed to follow a moist-adiabatic temperature gradient, and the altitude at which this adiabat first intersects the skin temperature inferred from the outgoing longwave radiation is taken as the tropopause. Above the tropopause, the atmosphere is held isothermal at this skin temperature. As initial guesses for the P–T profile, we adopt two limiting cases: (i) a fully isothermal atmosphere at the prescribed surface temperature, and (ii) an atmosphere in which the temperature gradient throughout the column follows the moist-adiabatic lapse rate. We have verified that the converged radiative–convective solution is independent of this choice, with both initial profiles relaxing to the same equilibrium state.

Beyond the fiducial 1~bar configuration, we also explore a grid of radiative–convective equilibrium solutions in the plane of H$_2$ surface pressure and planetary Bond albedo. These calculations are used to identify the combinations of pressure and albedo for which runaway-greenhouse
conditions are avoided, thereby delineating a habitable regime in pressure–albedo space; in this context, we classify
solutions as runaway or non-runaway according to the outgoing-longwave-radiation criterion described in
Section~\ref{sec:climate_stability}.

In this context, we further track how the stratospheric H$_2$O mixing ratio responds to changes in surface pressure and albedo via their influence on the tropopause temperature and vertical water vapour distribution. The resulting H$_2$O abundances provide a physical reference for the stratospheric H$_2$O mixing ratios treated as free parameters in the photochemical model (Section~\ref{sec: PM}), and enable a consistent interpretation of our CO/CO$_2$ mass-balance experiments in terms of the underlying climate state.

\section{Results}
\subsection{Photochemistry of $\mathrm{H_2}$-dominated atmosphere}\label{sec:photochemi}
\subsubsection{Chemical reaction network}

\begin{figure}
 \centering
 \includegraphics[width=\columnwidth]{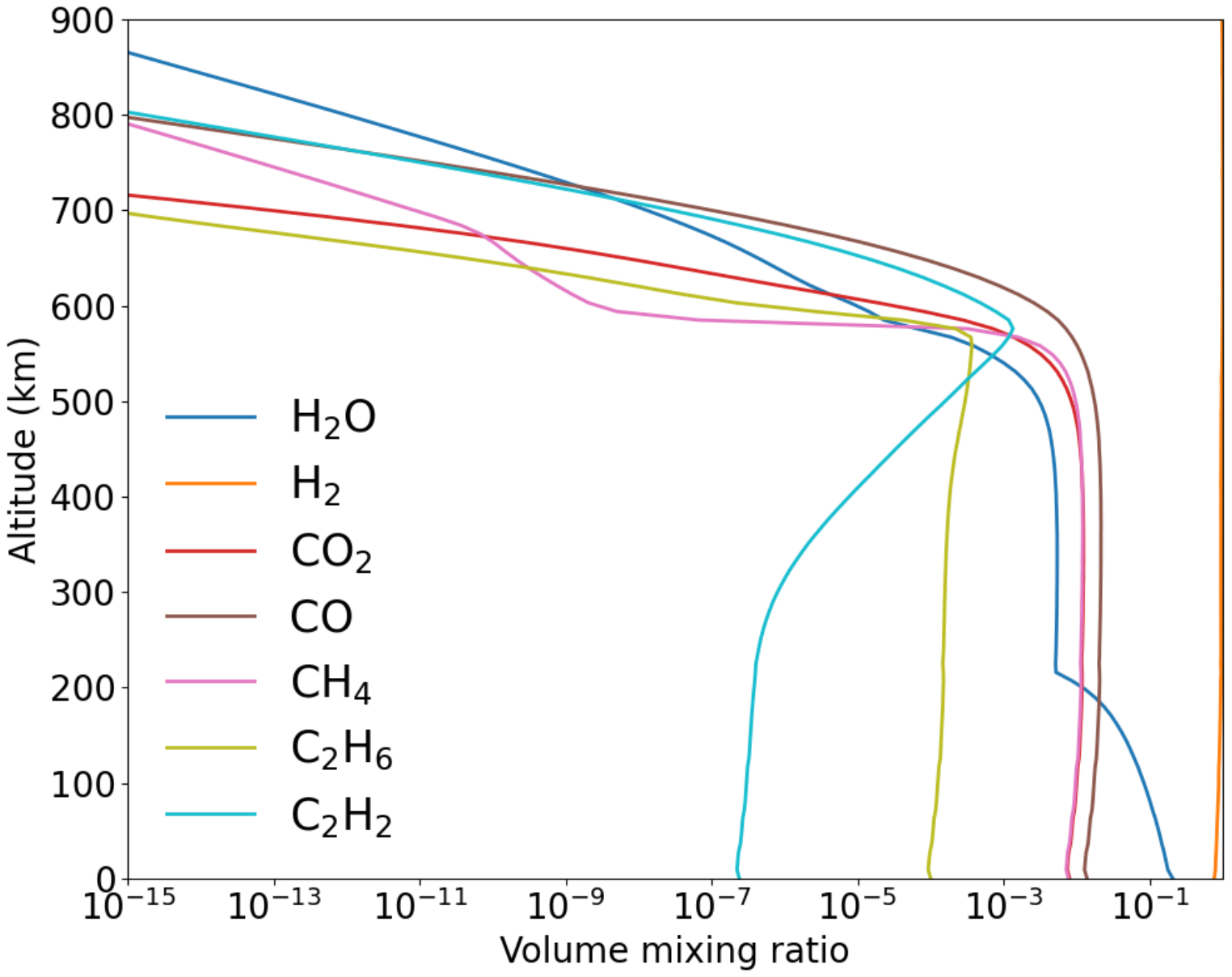}
 \caption{Altitude distribution of mixing ratios for the CO$_2$-bearing reference case obtained from our photochemical calculations.
In addition to the initial chemical species, enhanced concentrations of CO, C$_2$H$_6$, and C$_2$H$_2$ are observed,
with CO reaching mixing ratios of order 2\% and C$_2$H$_6$ and C$_2$H$_2$ reaching $\sim 10^{-3}$ near an altitude of
600~km. A sharp decrease in CH$_4$ occurs around the same altitude, where the C$_2$H$_2$ concentration is high.
}
\label{fig:mixing_ratios_co2_case}
\end{figure}

For the CO\textsubscript{2}-bearing reference case described in Section~\ref{sec: PM}, the number density distributions at steady state obtained from our photochemical calculations are shown in Figure~\ref{fig:mixing_ratios_co2_case}. In this case, the main photochemically produced species reach mixing ratios of approximately 2\% for CO and $\sim 10^{-3}$ for C\textsubscript{2}H\textsubscript{6} and C\textsubscript{2}H\textsubscript{2} at an altitude of about 600~km. 

\begin{figure*}
 \centering
 \includegraphics[width=\textwidth]{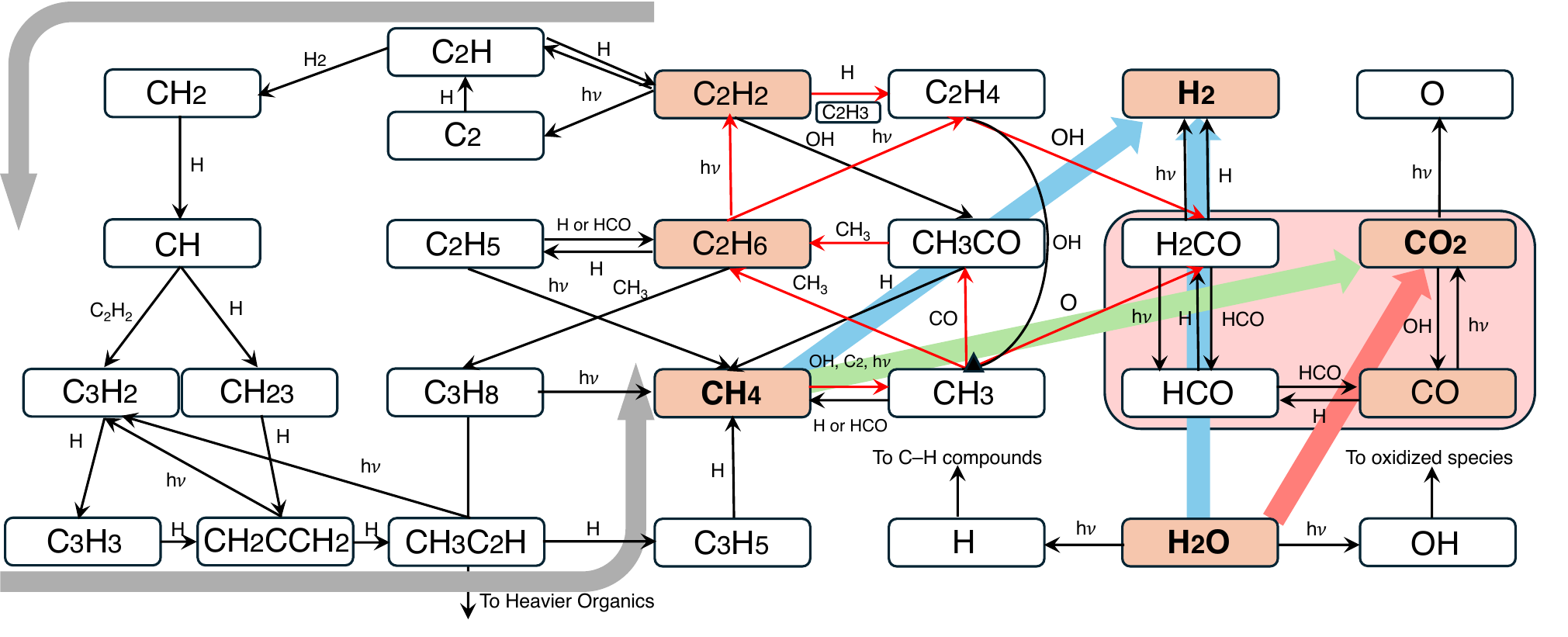}
 \caption{Diagram of the dominant chemical reaction pathways in our photochemical network.
    \textbf{Bold} text indicates the initial chemical species.
    Species shaded in light orange correspond to those plotted in Figure~\ref{fig:mixing_ratios_co2_case} and included in the absorption calculations for the transit spectrum shown in the upper panel of Figure~\ref{fig:transit_spectra}.
    Thick colored arrows represent the net transport of atoms: blue for H, green for C, and red for O.
    The thick gray curved arrow highlights the circulation loop of hydrocarbon (C--H) species.
    The thin light-red box on the right-hand side highlights the oxygen-bearing CO/CO$_2$ cycle into which a small fraction of carbon leaks from the hydrocarbon loop.
    Under the CO2-bearing reference case of this study, approximately 99\% of the carbon from CH$_4$ remains within the C--H circulation loop (thick gray arrow), while about 1\% is directed towards oxidation (red arrows), leading to its consumption.
    Arrows extending from C$_3$ species imply further polymerization leading to the formation of heavier organic hazes.}
\label{fig:reaction_network}
\end{figure*}

The production of CO is driven by the combined effects of methane oxidation and CO\textsubscript{2} photolysis, characteristic of an atmosphere where CH\textsubscript{4}, CO\textsubscript{2}, H\textsubscript{2}O, and H\textsubscript{2} coexist. As illustrated by the narrow bright-red arrows in Figure~\ref{fig:reaction_network}, carbon originating from CH\textsubscript{4} is transferred into the oxygen-bearing cycle primarily via the formation of H\textsubscript{2}CO in the thin light-red box on the right-hand side of Figure~\ref{fig:reaction_network}. Key reactions driving this transfer include the oxidation of methyl radicals by atomic oxygen (derived from CO\textsubscript{2} photolysis) and the oxidation of ethylene by hydroxyl radicals (derived largely from H\textsubscript{2}O):
\begin{align}
\text{CO\textsubscript{2}} + h\nu &\rightarrow \text{CO} + \text{O}, \label{eq:co2_phot} \\
\text{H\textsubscript{2}O} + h\nu &\rightarrow \text{H} + \text{OH}, \label{eq:h2o_phot} \\
\text{CH\textsubscript{3}} + \text{O} &\rightarrow \text{H\textsubscript{2}CO} + \text{H}, \label{eq:ch3_o_h2co} \\
\text{C\textsubscript{2}H\textsubscript{4}} + \text{OH} &\rightarrow \text{H\textsubscript{2}CO} + \text{CH\textsubscript{3}}. \label{eq:c2h4_oh_h2co}
\end{align}%
Subsequently, the photodissociation of H$_2$CO produces HCO, which is efficiently converted to CO via HCO self-reaction. The CO abundance is regulated primarily by CO-producing reactions involving HCO. For example,
\begin{align}
\text{HCO} + \text{HCO} &\rightarrow \text{H}_2\text{CO} + \text{CO}, \label{eq:hco_hco_new}
\end{align}%
and the consumption reaction is driven by the three-body association of H and CO,
\begin{align}
\text{H} + \text{CO} + \text{M} &\rightarrow \text{HCO} + \text{M}, \label{eq:h_co_m_new}
\end{align}
which sustains the CO mixing ratio.

\begin{figure}
 \centering
 \includegraphics[width=\columnwidth]{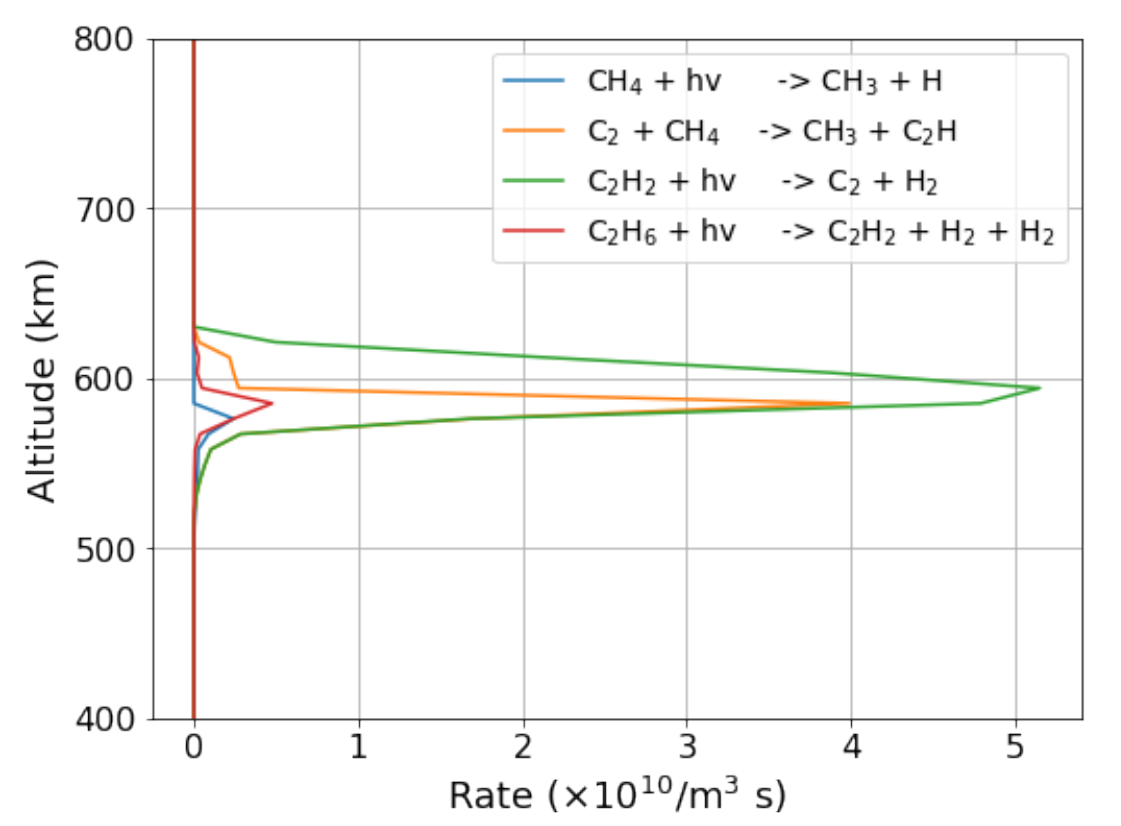}
 \caption{Graph showing the reaction rates of major CH\textsubscript{4} decomposition reactions in the region of CH\textsubscript{4} concentration change. Reactions involving C\textsubscript{2} radicals produced by the photodissociation of C\textsubscript{2}H\textsubscript{6} are dominant.}
\label{fig:ch4_decomposition_rates}
\end{figure}

As shown in Figure~\ref{fig:mixing_ratios_co2_case}, a sharp decrease in CH\textsubscript{4} number density is observed around 600~km. This decrease is caused not only by the photodissociation of CH\textsubscript{4} itself, but also by the enhanced CH\textsubscript{4} destruction triggered by the production of C\textsubscript{2}H\textsubscript{6} and C\textsubscript{2}H\textsubscript{2} and the subsequent reactions involving C\textsubscript{2}-containing radicals (Figure~\ref{fig:ch4_decomposition_rates}):
\begin{align}
\text{C\textsubscript{2}H\textsubscript{2}} + h\nu &\rightarrow \text{C\textsubscript{2}} + \text{H\textsubscript{2}}, \label{eq:c2h2_phot} \\
\text{C\textsubscript{2}} + \text{CH\textsubscript{4}} &\rightarrow \text{CH\textsubscript{3}} + \text{C\textsubscript{2}H}. \label{eq:c2_ch4}
\end{align}%
These C\textsubscript{2} radicals significantly enhance the CH\textsubscript{4} destruction efficiency at high altitudes, leading to the formation of the pronounced CH\textsubscript{4} depletion layer.

In the lower atmosphere (altitudes $\lesssim 400$ km), C$_2$H$_6$ is produced through the following reaction network:
\begin{align}
\text{H}_2\text{O} + h\nu &\rightarrow \text{H} + \text{OH}, \label{eq:7a_again} \\
\text{CH}_4 + \text{H} &\rightarrow \text{CH}_3 + \text{H}_2, \label{eq:7b_again} \\
\text{CH}_4 + \text{OH} &\rightarrow \text{CH}_3 + \text{H}_2\text{O}, \label{eq:7c_again} \\
\text{CO}_2 + h\nu &\rightarrow \text{CO} + \text{O}, \label{eq:7d_again} \\
\text{CO} + \text{CH}_3 + \text{M} &\rightarrow \text{CH}_3\text{CO} + \text{M}, \label{eq:7e_again} \\
\text{CH}_3\text{CO} + \text{CH}_3 &\rightarrow \text{CO} + \text{C}_2\text{H}_6. \label{eq:7_again}
\end{align}%
An alternative pathway involving HCO, also occurring below 400~km, is given by
\begin{align}
\text{H} + \text{CO} + \text{M} &\rightarrow \text{HCO}, \label{eq:8a_again} \\
\text{C\textsubscript{2}H\textsubscript{6}} + \text{H} &\rightarrow \text{C\textsubscript{2}H\textsubscript{5}} + \text{H\textsubscript{2}}, \label{eq:8b_again} \\
\text{C\textsubscript{2}H\textsubscript{6}} + \text{OH} &\rightarrow \text{C\textsubscript{2}H\textsubscript{5}} + \text{H\textsubscript{2}O}, \label{eq:8c_again} \\
\text{C\textsubscript{2}H\textsubscript{5}} + \text{HCO} &\rightarrow \text{C\textsubscript{2}H\textsubscript{6}} + \text{CO}. \label{eq:8_again}
\end{align}%
These pathways supply C\textsubscript{2}H\textsubscript{6} in the lower atmosphere via three-body reactions involving CO produced by CO\textsubscript{2} photolysis. Above 400~km, C\textsubscript{2}H\textsubscript{6} is mainly formed by
\begin{align}
\text{H} + \text{C\textsubscript{2}H\textsubscript{5}} + \text{M} &\rightarrow \text{C\textsubscript{2}H\textsubscript{6}} + \text{M}, \label{eq:9_again} \\
2\text{CH\textsubscript{3}} + \text{M} &\rightarrow \text{C\textsubscript{2}H\textsubscript{6}} + \text{M}, \label{eq:10_again}
\end{align}%
while at altitudes around 600~km it is in quasi-equilibrium due to photodissociation and radical reactions, in particular
\begin{align}
\text{C\textsubscript{2}H\textsubscript{6}} + h\nu &\rightarrow \text{C\textsubscript{2}H\textsubscript{2}} + \text{H\textsubscript{2}} + \text{H\textsubscript{2}}, \label{eq:11_again} \\
\text{C\textsubscript{2}H\textsubscript{6}} + h\nu &\rightarrow \text{C\textsubscript{2}H\textsubscript{4}} + \text{H} + \text{H}, \label{eq:12_again} \\
\text{C\textsubscript{2}H\textsubscript{6}} + h\nu &\rightarrow 2\text{CH\textsubscript{3}}, \label{eq:13_again} \\
\text{C\textsubscript{2}H\textsubscript{6}} + \text{H} &\rightarrow \text{C\textsubscript{2}H\textsubscript{5}} + \text{H\textsubscript{2}}, \label{eq:14_again} \\
\text{C\textsubscript{2}H\textsubscript{6}} + \text{OH} &\rightarrow \text{C\textsubscript{2}H\textsubscript{5}} + \text{H\textsubscript{2}O}. \label{eq:15_again}
\end{align}%
The increase in C\textsubscript{2}H\textsubscript{2} around 600~km is mainly supplied by reaction~(\ref{eq:11_again}), and the resulting C\textsubscript{2}H\textsubscript{2} is photodissociated by reaction~(\ref{eq:c2h2_phot}) to produce C\textsubscript{2} radicals. These C\textsubscript{2} radicals then react with CH\textsubscript{4} via reaction~(\ref{eq:c2_ch4}), efficiently decomposing CH\textsubscript{4} and producing CH\textsubscript{3} and C\textsubscript{2}H.

Through this sequence C\textsubscript{2}H\textsubscript{6} $\rightarrow$ C\textsubscript{2}H\textsubscript{2} $\rightarrow$ C\textsubscript{2}, the C\textsubscript{2}-radical-mediated CH\textsubscript{4} destruction channel is strongly enhanced at high altitudes. As a result, the efficiency of CH\textsubscript{4} destruction increases significantly near 600~km, producing the sharp CH\textsubscript{4} depletion layer seen in Figure~\ref{fig:mixing_ratios_co2_case}.

From the globally integrated flux analysis for the CO\textsubscript{2}-bearing reference case (Section~\ref{sec:CH4_mass_balance}), we find that, for a 1~bar H\textsubscript{2} envelope with a CH\textsubscript{4} mixing ratio of 1\%, CH\textsubscript{4} is irreversibly converted into CO and CO\textsubscript{2} at a net rate of $\sim 5.5 \times 10^{3}$~kg~s\textsuperscript{$-1$}. This corresponds to a characteristic photochemical lifetime of CH\textsubscript{4} of $\sim 1.2 \times 10^{7}$~yr for a 1\% CH\textsubscript{4} abundance in a 1~bar H\textsubscript{2} atmosphere, which is much shorter than the estimated age of the system (several Gyr). Thus, even in the CO\textsubscript{2}-bearing case, the reaction network described above, including the C\textsubscript{2}H\textsubscript{6}–C\textsubscript{2}H\textsubscript{2}–C\textsubscript{2} chain, continuously converts CH\textsubscript{4} into CO/CO\textsubscript{2} on relatively short timescales. At the same time, within the photochemical model, H\textsubscript{2} itself is produced at a small net rate rather than destroyed, so that its long-term abundance is ultimately controlled by atmospheric escape and interior outgassing, as discussed in Section~\ref{sec:CH4,H2}.

Under the fiducial conditions of this study, the vast majority of carbon originating from CH\textsubscript{4} remains within the closed hydrocarbon (C–H) loop on the left-hand side of Figure~\ref{fig:reaction_network}, while only a small fraction leaks into the oxidised CO/CO\textsubscript{2} branch in the thin light-red box on the right-hand side. This organic circulation continuously feeds higher-order hydrocarbons such as C\textsubscript{3}H\textsubscript{8} and its derivatives, which in turn provide the precursors for further polymerisation into more complex organic material and haze particles. We return to the implications of this hydrocarbon loop for haze production and the planetary Bond albedo in Section~\ref{sec:haze_albedo} and Appendix~\ref{sec:appendix_haze}.

\subsubsection{$\mathrm{CO}$ and $\mathrm{CO_2}$ production and dependence on stratospheric $\mathrm{H_2O}$}
\label{sec:CO2_H2O_dependence}

\begin{figure*}
 \centering
 \includegraphics[width=\textwidth]{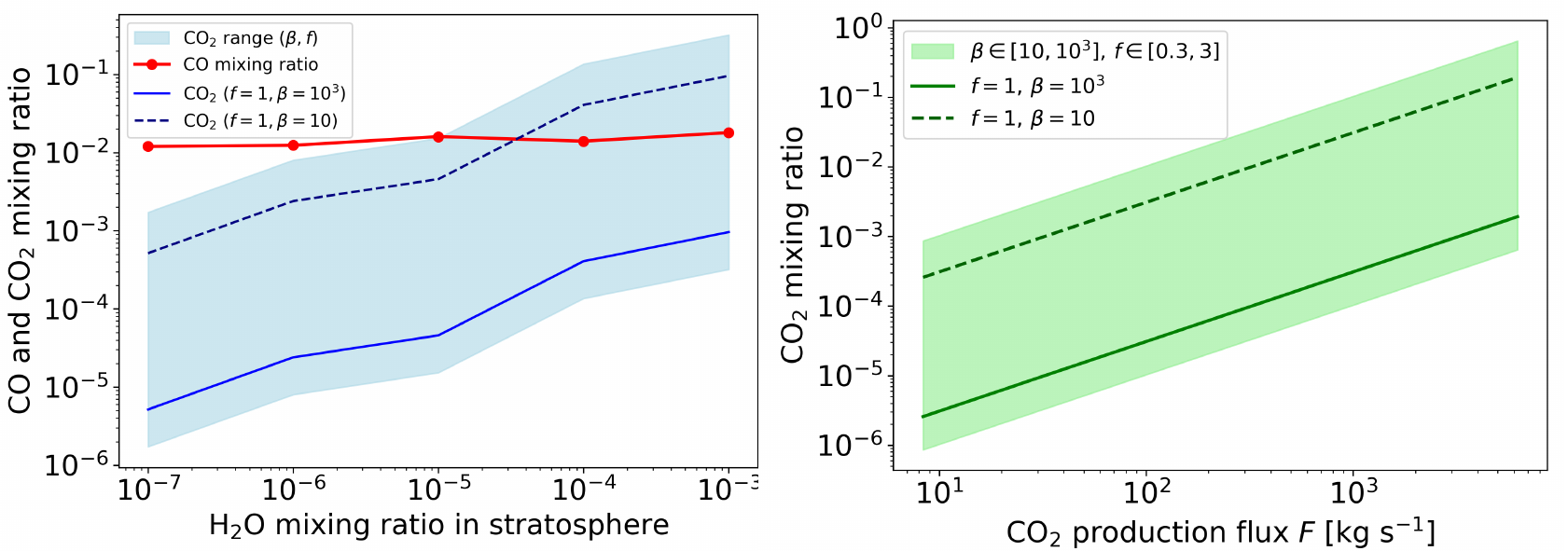}
 \caption{
Relationships between the prescribed stratospheric H$_2$O mixing ratio, the CO$_2$ mass flux, and the resulting atmospheric CO$_2$ mixing ratio in the CO$_2$-free experiments.
Left panel: CO$_2$ mixing ratio accumulated in the atmosphere over a timescale of 3~Gyr as a function of the stratospheric H$_2$O mixing ratio in the photochemical experiments lower-boundary-CO2-free experiments. Right panel: atmospheric CO$_2$ mixing ratio as a function of the vertically integrated CO$_2$ mass flux. In both panels, the shaded regions indicate the ranges obtained when varying the ocean parameters $\beta$ and $f$, while the solid and dashed curves correspond to alkaline and acidic ocean conditions, respectively. The stratospheric H$_2$O mixing ratio is evaluated at the cold trap near the tropopause.
 }
\label{fig:co2_h2o_massbalance}
\end{figure*}

In our photochemical experiments for H$_2$--CH$_4$--H$_2$O atmospheres, we find that, even when varying the lower-boundary CO$_2$ conditions and the prescribed stratospheric H$_2$O mixing ratio over the range $10^{-7}$--$10^{-3}$, the coupled H--C--O reaction network robustly drives CO toward mixing ratios of order 1--2\% as long as these three components coexist (Figure~\ref{fig:co2_h2o_massbalance}). As we show in Section~\ref{sec:climate_stability}, this H$_2$O range broadly overlaps with that expected for non--runaway, Hycean-like climates on K2-18b.

We also find that the net CH$_4$ consumption flux in the CO$_2$-free experiments is comparable in magnitude to that obtained in the CO$_2$-bearing reference case. This behaviour arises because, as illustrated in Figure~\ref{fig:reaction_network}, most of the carbon from photodissociated CH$_4$ is recycled within the hydrocarbon (C--H) loop on the left-hand side of the network. 

In the experiments lower-boundary-CO2-free experiments, the evolution of atmospheric CO$_2$ is controlled by a small subset of reactions in the H--C--O network. The net production of CO$_2$ from reduced carbon proceeds mainly via
\begin{align}
    \mathrm{CO} + \mathrm{OH} &\rightarrow \mathrm{CO_2} + \mathrm{H},
    \label{eq:CO_OH_CO2}
\end{align}
so that the CO$_2$ production flux is determined primarily by the availability of OH and CO and is highly sensitive to the prescribed stratospheric H$_2$O mixing ratio. Quantitatively, the vertically integrated CO$_2$ production fluxes for representative stratospheric H$_2$O mixing ratios are in Table~\ref{tab:CO2_flux}.
\begin{table}[t]
    \centering
    \caption{Vertically integrated CO$_2$ production flux as a function of the prescribed stratospheric H$_2$O mixing ratio in the CO$_2$-free experiments.}
    \label{tab:CO2_flux}
    \begin{tabular}{cc}
        \hline\hline
        Stratospheric H$_2$O mixing ratio & CO$_2$ production flux [kg s$^{-1}$] \\
        \hline
        $10^{-3}$ & $9.45 \times 10^{2}$ \\
        $10^{-4}$ & $4.05 \times 10^{1}$ \\
        $10^{-5}$ & $1.92 \times 10^{0}$ \\
        $10^{-6}$ & $3.77 \times 10^{-1}$ \\
        $10^{-7}$ & $2.66 \times 10^{-1}$ \\
        \hline
    \end{tabular}
\end{table}

To relate the photochemically produced CO$_2$ to an atmospheric CO$_2$ mixing ratio under Hycean conditions, we combine the CO$_2$ production flux with a simplified mass-balance framework that includes dissolution into a global ocean and oceanic carbon buffering (see Appendix~\ref{sec:appendix_CO2mass} for details). In this framework, the oceanic uptake of CO$_2$ is characterised by two parameters. The first is the DIC enhancement factor $\beta$, defined as the ratio of the total dissolved inorganic carbon (DIC) to dissolved molecular CO$_2$(aq). The second is the ocean-depth parameter $f$, which scales the ocean volume relative to a reference Hycean configuration. Larger values of $\beta$ and $f$ correspond to a greater capacity of the ocean to store oxidised carbon at fixed atmospheric CO$_2$ partial pressure. The factor $\beta$ effectively encapsulates the dependence of carbonate speciation on ocean pH: for present-day Earth-like mildly alkaline conditions (pH~$\sim 8$), $\beta$ is typically of order $10^2$, while more strongly alkaline oceans correspond to $\beta$ values of order $10^2$--$10^3$.


The left panel of Figure~\ref{fig:co2_h2o_massbalance} summarises, for the CO$_2$-free photochemical experiments, the atmospheric CO$_2$ mixing ratio that can be maintained over gigayear timescales for a given stratospheric H$_2$O abundance when ocean uptake is taken into account. As the stratospheric H$_2$O abundance is reduced, the supply of OH radicals that drive reaction~(\ref{eq:CO_OH_CO2}) decreases, and the resulting CO$_2$ production flux, as well as the inferred steady-state atmospheric CO$_2$ mixing ratio, both decrease systematically. The right panel of Figure~\ref{fig:co2_h2o_massbalance} shows, for different choices of $\beta$ and $f$, the relationship between the vertically integrated CO$_2$ mass flux and the corresponding atmospheric CO$_2$ mixing ratio. The connection between these results and the surface temperature and pressure regime is discussed further in Section~\ref{subsec:H2O_CO2_Hycean}.

\subsection{Transit Spectrum}\label{subsec:transit_spectrum}

\subsubsection{Comparison between photochemical and uniform profiles}\label{subsubsec:transit_profiles}

\begin{figure*}
 \centering
 \includegraphics[width=\textwidth]{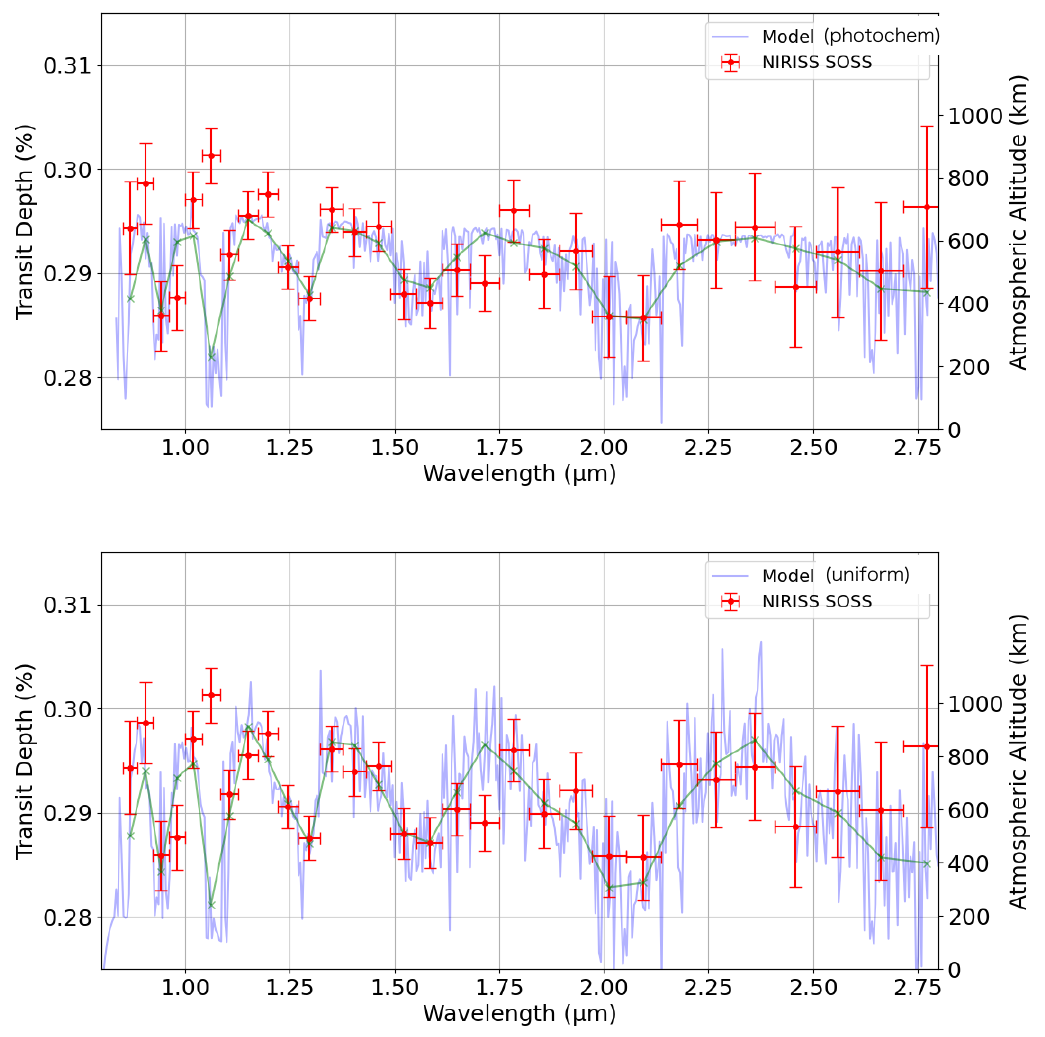}
 \caption{Transit spectra of K2-18b in the NIRISS SOSS wavelength
    range (0.8–2.8~$\micron$). The figure compares model
    spectra (light purple lines) with observed data points
    including error bars. In the upper panel, we use the
    number density profiles of H$_2$, CH$_4$, CO, CO$_2$,
    C$_2$H$_2$, and C$_2$H$_6$ obtained from our
    photochemical calculations. In the lower panel, we
    instead assume vertically uniform number densities for
    all species. Following \citet{Madhusudhan2023}, the
    mixing ratios of CH$_4$ and CO$_2$ are each set to 1\%,
    while H$_2$O, CO, DMS, and H$_2$O are each set to
    10$^{-5}$. The green dashed lines show the model spectra
    averaged over the observational bin widths. For the
    NIRISS SOSS wavelength range, the upper-panel model
    yields $\chi^2 = 3.154$, whereas the lower-panel model
    yields $\chi^2 = 4.257$. Other model parameters are
    listed in Table \ref{tab3}.}
\label{fig:transit_spectra}
\end{figure*}

Figure~\ref{fig:transit_spectra} compares the model transmission spectra
computed for the NIRISS SOSS data set of
\citet{Madhusudhan2023} using (i) the altitude-dependent mixing-ratio
profiles from our photochemical calculations and (ii) vertically uniform
mixing ratios following \citet{Madhusudhan2023}. In both cases we adopt
the Hycean-like temperature--pressure structure shown in
Figure~\ref{fig:pt_profile} and the same aerosol parameterization
described in Section~2.2, and we allow the reference planet-to-star
radius ratio $R_p/R_s$ to vary only within the observational
uncertainty (Table~\ref{tab1}).

For the \citet{Madhusudhan2023}'s reduction, the uniform-profile case yields
$\chi^2 = 4.257$ for 24 degrees of freedom in the NIRISS SOSS range,
whereas the photochemical profiles give $\chi^2 = 3.154$, indicating a
moderate but systematic improvement. For the independent reductions
\texttt{exoTEDRF} and \texttt{FIREFLy}, we likewise find that photochemical profiles
provide equal or better fits in the CH$_4$-dominated region. At a
representative resolution of $R\approx 25$, the reduced $\chi^2$ values
for the NIRISS band are summarised in Table~\ref{tab:chi2_profiles},
showing that our photochemical profiles produce lower
$\chi^2_\nu$ than the uniform profiles in both cases.

\begin{deluxetable}{lcc}
\tablecaption{Reduced $\chi^2$ values in the NIRISS SOSS band
($0.8$--$2.8~\micron$) for different vertical profiles and
reductions.\label{tab:chi2_profiles}}
\tablehead{
\colhead{Reduction} & \colhead{Uniform profile} & \colhead{Our photochemical profile}
}
\startdata
Madhusudhan et al. (2023)   ($R\approx 25$) & 4.257 & 3.154 \\
\texttt{exoTEDRF}  ($R\approx 25$) & 2.501 & 1.949 \\
\texttt{FIREFLy}   ($R\approx 25$) & 2.379 & 1.952 \\
\enddata
\end{deluxetable}

In the NIRISS wavelength range ($0.8$--$2.8~\micron$), where CH$_4$
features dominate, the spectrum based on vertically uniform mixing
ratios closely follows the wavelength dependence of the CH$_4$
absorption coefficients (Figure~\ref{fig:opacity_spectra}), producing
pronounced peaks and troughs that reflect the line strengths. In
contrast, the spectrum obtained with the photochemical profiles exhibits
a clear flattening across the CH$_4$ bands. This difference does not
arise simply from increasing the CH$_4$ abundance at all altitudes. In
the uniform case, raising the CH$_4$ mixing ratio primarily increases
the optical depth and slightly reduces the scale height through an
increase in mean molecular weight, leading to moderate changes in band
amplitudes but not to the emergence of an extended plateau.

In the photochemical calculations, CH$_4$ is strongly depleted at high
altitudes by photodissociation (Figure~\ref{fig:mixing_ratios_co2_case}),
so that, over a broad wavelength range where CH$_4$ opacity dominates,
the effective transit radius is determined by nearly the same altitude
just below the CH$_4$ depletion layer. This ``photochemical saturation''
produces a flat-topped spectral structure that mimics, in the vertical
direction, a common cutoff height for CH$_4$-dominated features. 

\subsubsection{Offset constraints from the $\mathrm{CH_4}$ band}\label{subsubsec:offset_results}

\begin{figure*}
\centering
\includegraphics[width=\textwidth]{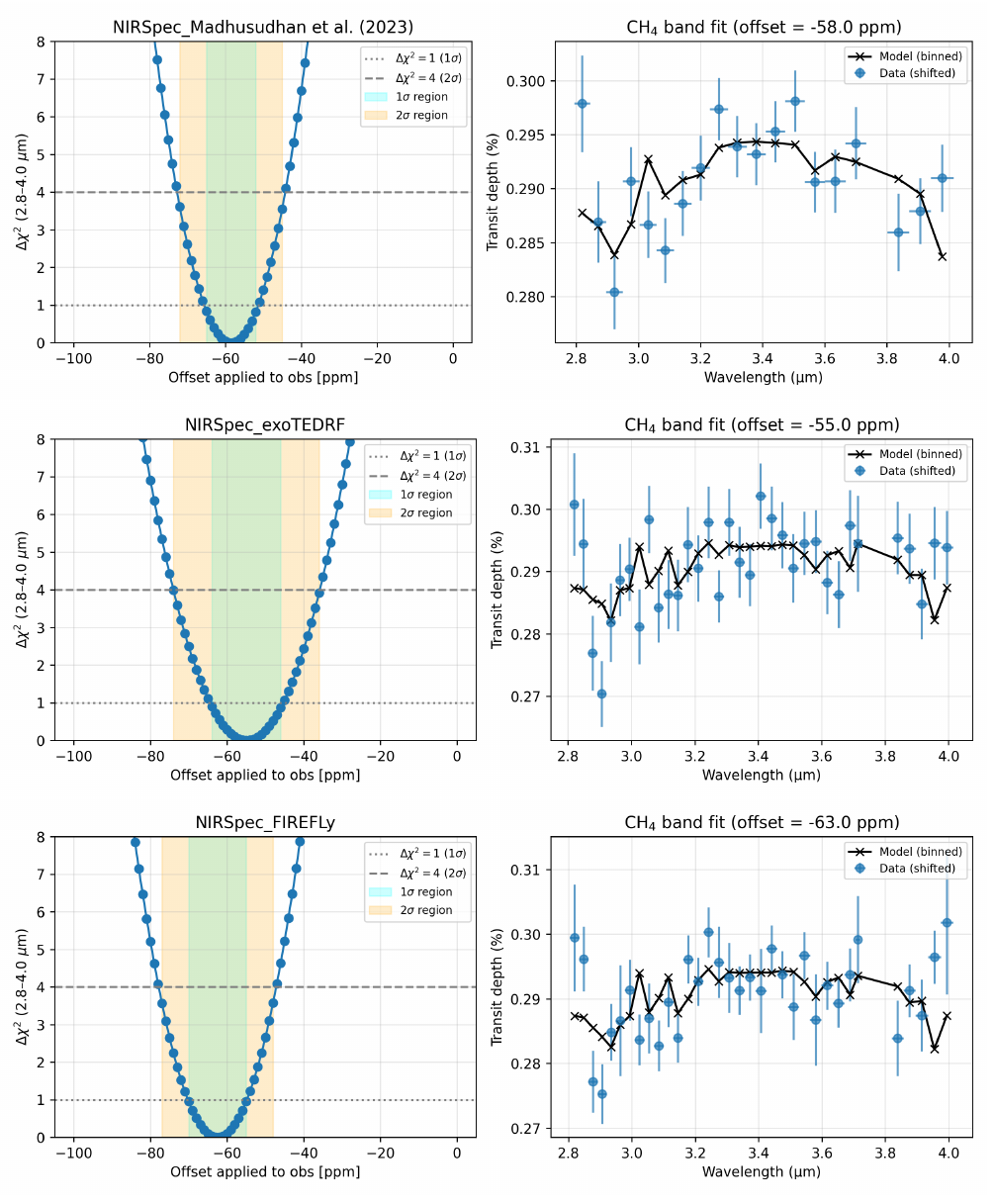}
\caption{
Left: Posterior distributions of the wavelength-independent
offset $\Delta_{\rm off}$ for each reduction, derived from the
CH$_4$-dominated $2.8$–$4.0~\micron$ band using
Equation~(\ref{eq:post_offset}). Shaded regions indicate
the $1\sigma$ and $2\sigma$ credible intervals listed in
Table~\ref{tab:offsets}. Right: Comparison between the
best-fit Hycean model spectra (solid lines) and the
offset-applied data (points with error bars) in the
$2.8$–$4.0~\micron$ range for the
\citet{Madhusudhan2023} reduction, \texttt{exoTEDRF},
and \texttt{FIREFLy}. 
}
\label{fig:offset_panel}
\end{figure*}

Figure~\ref{fig:offset_panel} summarises the offset
constraints obtained from the CH$_4$ band. The left
panel shows, for each reduction, the probability
distribution $p(\Delta_{\rm off})$ defined by
Equation~(\ref{eq:post_offset}), together with the
$1\sigma$ and $2\sigma$ credible intervals. The right
panels compare the best-fit Hycean model spectra (solid
lines) and the offset-applied data (points with error
bars) in the $2.8$–$4.0~\micron$ range for the
\citet{Madhusudhan2023} reduction, \texttt{exoTEDRF},
and \texttt{FIREFLy}.

\begin{deluxetable}{lcc}
\tablecaption{Offsets constrained from the CH$_4$ band
($2.8$–$4.0~\micron$). For the
\citet{Madhusudhan2023} reduction we use $R\approx 55$,
while \texttt{exoTEDRF} and \texttt{FIREFLy} are
analysed at $R\approx 100$. For \texttt{Eureka!  Reduction A} and
\texttt{Eureka! Reduction B} we do not introduce an additional
offset parameter and set $\Delta_{\rm off}=0$; see
Section~\ref{subsubsec:method_offsets}.\label{tab:offsets}}
\tablehead{
\colhead{Reduction} & \colhead{Resolution} &
\colhead{$\Delta_{\rm off}$ (1$\sigma$) [ppm]}
}
\startdata
Madhusudhan et al. (2023) & $R\approx 55$  & $[-65,-52]$ \\
\texttt{exoTEDRF}         & $R\approx 100$ & $[-64,-46]$ \\
\texttt{FIREFLy}          & $R\approx 100$ & $[-70,-55]$ \\
\texttt{Eureka! Reduction A}         & $R\approx 100$ & fixed at $0$ \\
\texttt{Eureka! Reduction B}         & $R\approx 100$ & fixed at $0$ \\
\enddata
\end{deluxetable}

For the \citet{Madhusudhan2023} reduction, the
$1\sigma$ range defined by $\Delta\chi^2_{\rm CH4}<1$ is
$\Delta_{\rm off}\simeq[-65,-52]$~ppm. For the
$R\sim 100$ analyses of \textsc{exoTEDRF} and
\texttt{FIREFLy}, we obtain
$\Delta_{\rm off}\simeq[-64,-46]$~ppm and
$\Delta_{\rm off}\simeq[-70,-55]$~ppm, respectively.
For \texttt{Eureka! Reduction A} and \texttt{Eureka! Reduction B}, the lack of
simultaneous NIRISS SOSS coverage means that their
absolute normalisation was already chosen in the
original retrievals by matching the CH$_4$ features
\citep{Schmidt2025}; we therefore do not allow an
additional offset and keep $\Delta_{\rm off}=0$ for
these reductions. By construction, the ranges listed
in Table~\ref{tab:offsets} correspond to offsets that
do not significantly degrade the fit to the CH$_4$
features, and therefore define the range of
normalizations for which the characteristic CH$_4$
structure remains compatible with the observations.

\subsubsection{$\mathrm{CO}$ and $\mathrm{CO_2}$ constraints from the 4–5~$\micron$ band}\label{subsec:CO_results}

\begin{figure*}
\centering
\includegraphics[width=\textwidth]{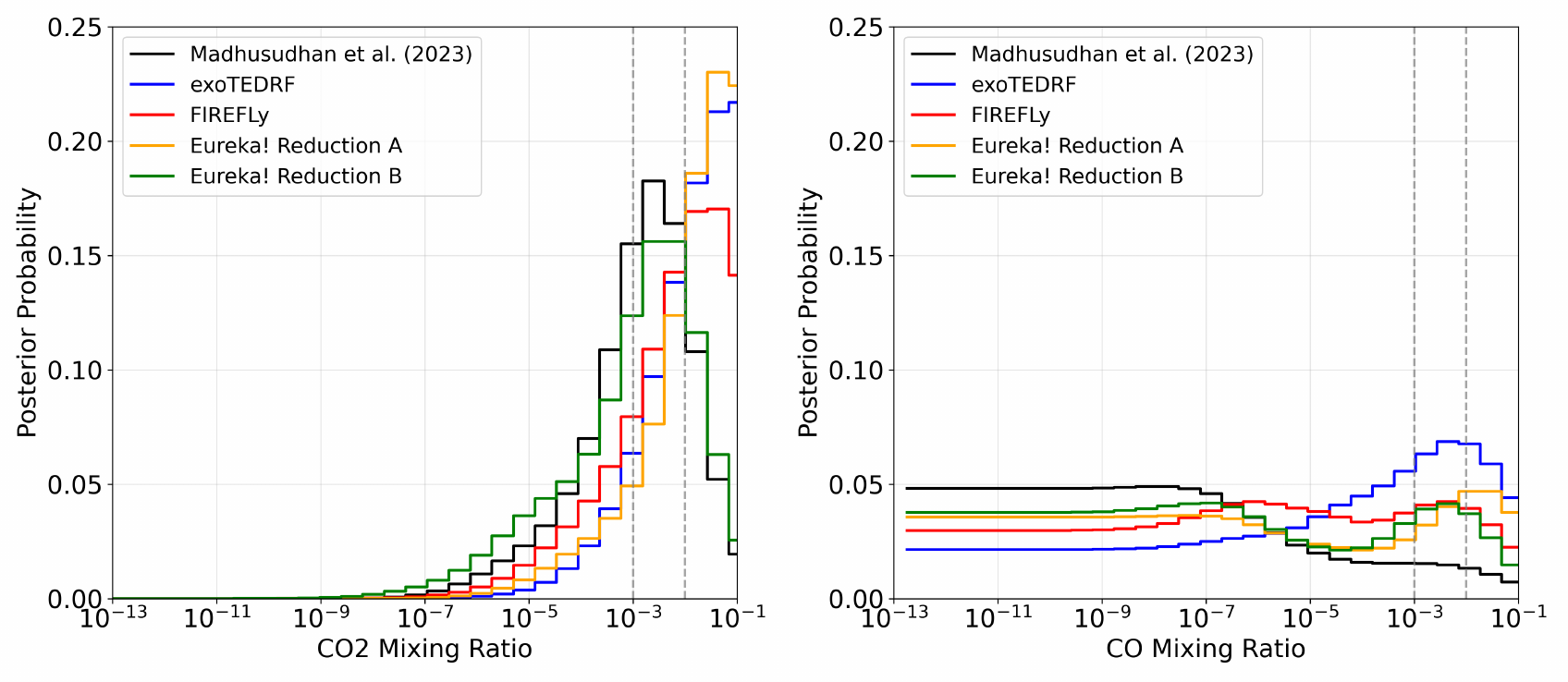}
\caption{
Posterior probability distributions for the CO (left) and CO$_2$
(right) volume mixing ratios, $p(X_{\rm CO})$ and $p(X_{\rm CO_2})$, for the
five reductions considered in this work. Each curve is plotted as a step
histogram in $\log_{10}$ mixing ratio and represents a probability density
in $u=\log_{10} X$ normalised such that $\int p(u)\,{\rm d}u = 1$. These
densities are obtained from the discrete posteriors on the log-spaced grid
by dividing by the bin width in $u$. All panels share the
same vertical scale to allow a direct visual comparison between CO and
CO$_2$. While most reductions do not strongly require high CO abundances,
moderately high CO$_2$ levels remain broadly allowed, with \texttt{exoTEDRF} and
\texttt{Eureka! Reduction~A }favouring the most CO$_2$-rich solutions. Quantitative
summary statistics are given in Table~\ref{tab:CO_stats}.
}
\label{fig:pCO_panels}
\end{figure*}

Figure~\ref{fig:pCO_panels} summarises the resulting CO and CO$_2$ posteriors for the five
reductions. The left panel shows $p(X_{\rm CO})$ and the right panel
$p(X_{\rm CO_2})$, each plotted as a step histogram in logarithmic mixing
ratio and normalised to unit area. For this figure, we convert the discrete
posteriors on the log-spaced grid into probability densities in
$u = \log_{10} X$ by dividing by the bin width in $u$ and renormalising so
that $\int p(u)\,{\rm d}u = 1$.
Qualitatively, most reductions allow relatively low CO abundances while still
permitting moderately high CO$_2$ levels, with \texttt{exoTEDRF} showing the strongest
preference for CO and \texttt{Eureka!} Reduction~A displaying the most CO$_2$-rich
solutions.

In the $R\approx 100$ reductions, the CO posteriors for
\texttt{exoTEDRF} and \texttt{Eureka! Reduction A} exhibit clear
peaks at $X_{\rm CO}\gtrsim 10^{-3}$, with best-fit values
$X_{\rm CO,best}\simeq 4.0\times 10^{-3}$ and
$2.7\times 10^{-2}$, respectively (Table~\ref{tab:CO_stats}).
At the same time, the probability that $X_{\rm CO}$
lies below $10^{-3}$ remains substantial in all
$R\approx 100$ analyses: even for the most CO-rich
case, \texttt{exoTEDRF}, we find $P(X_{\rm CO}\ge 10^{-3})\simeq 0.33$,
so that $P(X_{\rm CO}<10^{-3})\simeq 0.67$, while for
\texttt{FIREFLy}, \texttt{Eureka! Reduction A}, and
\texttt{Eureka! Reduction B} the corresponding
probabilities are
$P(X_{\rm CO}<10^{-3})\simeq 0.81$, $0.77$, and $0.84$,
respectively (see Table~\ref{tab:CO_stats}). Similarly,
the probability of very high CO abundances is small:
$P(X_{\rm CO}\ge 10^{-2})$ remains at
$\simeq 0.20$ (\texttt{exoTEDRF}),
$0.11$ (\texttt{FIREFLy}),
$0.16$ (\texttt{Eureka! Reduction A}), and
$0.085$ (\texttt{Eureka! Reduction B}).
Thus, while some $R\approx 100$ reductions favour
solutions with a peak in $p(X_{\rm CO})$ above $10^{-3}$,
the posteriors still assign relatively high probability to
CO mixing ratios below $10^{-3}$, and very CO-rich
($\gtrsim 10^{-2}$) atmospheres are never strongly required.

The quantitative constraints on CO and CO$_2$ are
summarised in Table~\ref{tab:CO_stats}.  For the
Madhusudhan et al. (2023) reduction analysed at
$R\approx 55$, the posterior for CO is strongly peaked
at very low mixing ratios and remains consistent with
$X_{\rm CO}\lesssim 10^{-3}$ at the $2\sigma$ level, while
CO$_2$ is only weakly constrained and allows mixing
ratios of order a few $\times 10^{-2}$. Among the
$R\approx 100$ reductions, \texttt{exoTEDRF} yields the
CO- and CO$_2$-richest solutions, favouring $X_{\rm CO}$
of order $10^{-3}$–$10^{-2}$ and $X_{\rm CO_2}$
approaching $10^{-1}$, whereas \texttt{FIREFLy} and
both \texttt{Eureka} reductions are all compatible with
CO mixing ratios below $10^{-3}$ while still permitting
CO$_2$ at $X_{\rm CO_2}\gtrsim 10^{-3}$ with probabilities
of order $0.5$–$0.8$. Taken together, these results indicate
that, once the CH$_4$-band-based offset constraints are
imposed, high CO abundances are not strongly required
by most reductions, whereas moderately high CO$_2$
abundances remain broadly allowed.

\begin{deluxetable*}{lcccccc}
\tablecaption{CO and CO$_2$ constraints from the 4–5~$\micron$
band, combining the CH$_4$-band offset priors with the
CO/CO$_2$-sensitive region. For the
\citet{Madhusudhan2023} reduction we use $R\approx 55$,
and for the other reductions we adopt $R\approx 100$.
\label{tab:CO_stats}}
\tablehead{
\colhead{Reduction} &
\colhead{Res.} &
\colhead{$X_{\rm CO,\,best}$} &
\colhead{$X_{\rm CO,\,2\sigma}$} &
\colhead{$P(X_{\rm CO}\ge 10^{-3})$} &
\colhead{$P(X_{\rm CO}\ge 10^{-2})$} \\
\colhead{} & \colhead{} &
\colhead{$X_{\rm CO_2,\,best}$} &
\colhead{$X_{\rm CO_2,\,2\sigma}$} &
\colhead{$P(X_{\rm CO_2}\ge 10^{-3})$} &
\colhead{$P(X_{\rm CO_2}\ge 10^{-2})$}
}
\startdata
Madhusudhan et al. (2023) &
$R\approx 55$ &
$6.4\times 10^{-9}$ &
$4.0\times 10^{-3}$ &
$0.065$ &
$0.035$ \\
& & $2.2\times 10^{-3}$ &
$3.9\times 10^{-2}$ &
$0.526$ &
$0.179$ \\
\texttt{exoTEDRF} &
$R\approx 100$ &
$4.0\times 10^{-3}$ &
$6.9\times 10^{-2}$ &
$0.331$ &
$0.199$ \\
& & $1.0\times 10^{-1}$ &
$1.0\times 10^{-1}$ &
$0.846$ &
$0.611$ \\
\texttt{FIREFLy} &
$R\approx 100$ &
$7.5\times 10^{-7}$ &
$2.7\times 10^{-2}$ &
$0.191$ &
$0.107$ \\
& & $3.9\times 10^{-2}$ &
$1.0\times 10^{-1}$ &
$0.732$ &
$0.481$ \\
\texttt{Eureka! Reduction A} &
$R\approx 100$ &
$2.7\times 10^{-2}$ &
$6.9\times 10^{-2}$ &
$0.228$ &
$0.155$ \\
& & $3.9\times 10^{-2}$ &
$1.0\times 10^{-1}$ &
$0.840$ &
$0.640$ \\
\texttt{Eureka! Reduction B} &
$R\approx 100$ &
$1.1\times 10^{-7}$ &
$2.7\times 10^{-2}$ &
$0.165$ &
$0.085$ \\
& & $2.2\times 10^{-3}$ &
$3.9\times 10^{-2}$ &
$0.517$ &
$0.205$ \\
\enddata
\end{deluxetable*}

\begin{deluxetable}{lll}
\tablecaption{Model parameters and their treatment in this work. For the CO and
CO$_2$ mixing ratios and the NIRSpec–NIRISS offset we adopt explicit
Bayesian priors as listed, while all other parameters are fixed to the
best-fit values given in Table~1 rather than being re-sampled in our
CO/CO$_2$ grid scan. \label{tab3}}
\tablehead{
\colhead{Parameter} & \colhead{Prior / treatment} & \colhead{Description}
}
\startdata
$\log(X_{\text{CO}_2})$ & $\mathcal{U}(-13, -1)$ & Mixing ratio of CO\textsubscript{2} \\
$\log(X_{\text{CO}})$ & $\mathcal{U}(-13, -1)$ & Mixing ratio of CO\\
$\delta_{\text{NIRspec}}/\text{ppm}$ & $\mathcal{U}(-100, 100)$ & NIRSpec dataset offset \\
\enddata
\end{deluxetable}

\subsection{Radiative–Convective Climate Structure and Its Dependence on Surface Pressure and Albedo} \label{sec:climate_stability}

\begin{figure}
 \centering
 \includegraphics[width=\columnwidth]{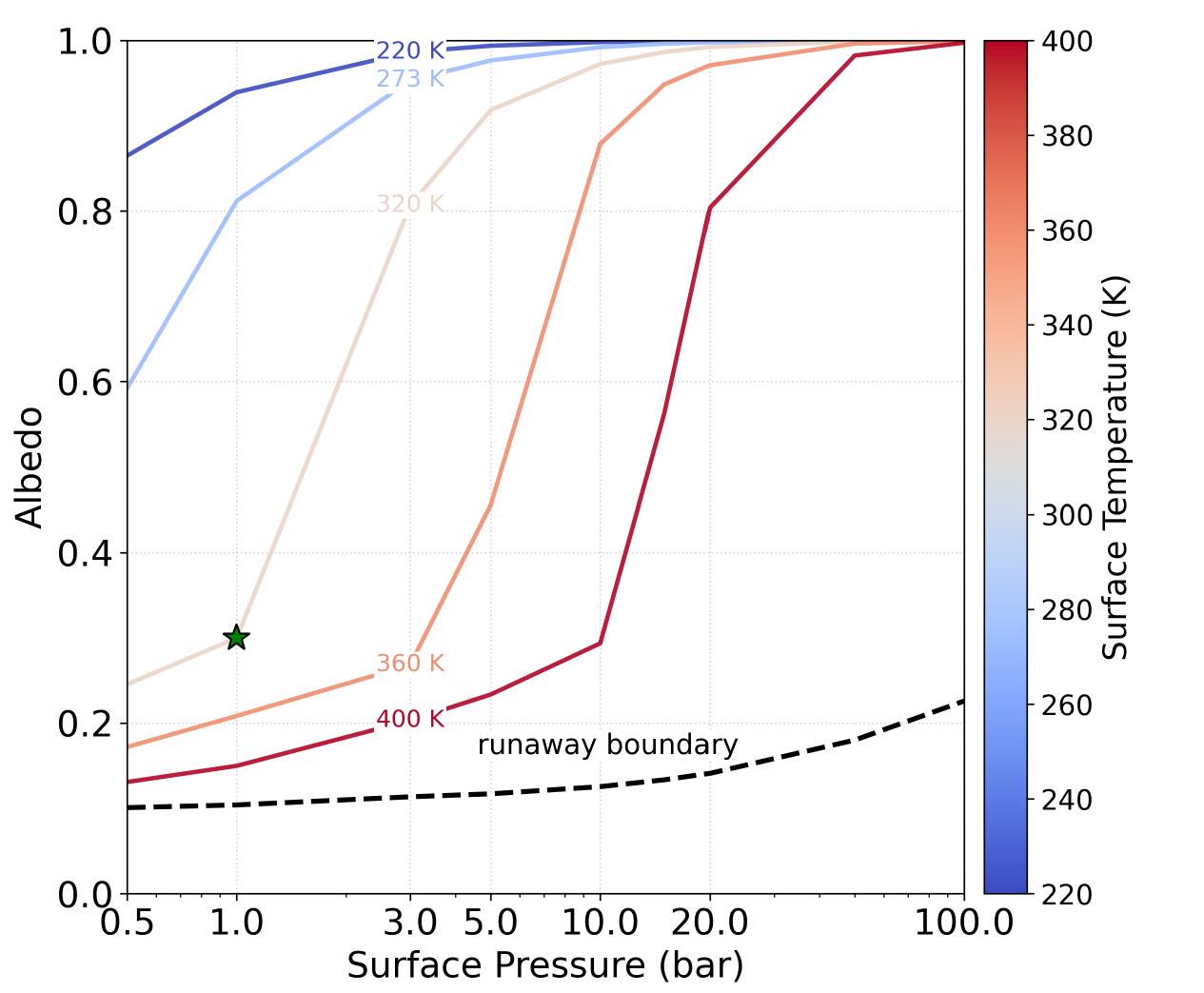}
 \caption{
Surface pressure--albedo diagram for H$_2$-rich atmospheres under the incident flux of K2-18b.
The horizontal axis shows the surface pressure, and the vertical axis shows the planetary Bond albedo.
Solid curves denote isotherms of surface temperature, colour-coded according to the scale bar on the right-hand side.
The dashed curve marks the approximate boundary beyond which no radiative--convective equilibrium with a subcritical liquid-water surface is obtained.
The green star indicates the surface condition (1~bar, 328~K) adopted in the fiducial photochemical calculations.
 }
 \label{fig:pressure_albedo}
\end{figure}

Our calculations reveal climatic constraints on maintaining a habitable climate under the Hycean assumption. Figure~\ref{fig:pressure_albedo} summarises the results of our radiative–convective calculations in the plane of H$_2$ surface pressure and planetary Bond albedo. For each combination of surface pressure and albedo, we obtain a radiative–convective equilibrium solution and compute the corresponding surface temperature. Solid curves show isotherms of $T_{\rm s}$, colour-coded according to the scale bar on the right-hand side, while the dashed curve marks the approximate runaway-greenhouse boundary beyond which a subcritical liquid-water surface can no longer be maintained. As the H$_2$ surface pressure increases along a given Bond albedo, the equilibrium surface temperature rises, and for sufficiently high pressures only very hot solutions remain. The fiducial Hycean configuration adopted in our photochemical calculations (a 1 bar H$_2$ envelope with a Bond albedo of $A_{\rm B}\sim 0.3$) lies in the region of Figure~\ref{fig:pressure_albedo} where moderately warm, non–runaway solutions with $T_{\rm s}\sim 320$–330 K are obtained, in good agreement with previous Hycean climate studies \citep[e.g.,][]{Wogan2024}.

Among the atmospheric constituents, we find that H$_2$O continuum absorption and H$_2$–H$_2$ collision-induced absorption provide the dominant contributions to the greenhouse effect. By contrast, the additional greenhouse forcing from the major photochemically produced species (CO, CH$_4$, CO$_2$, and C$_2$H$_6$) is comparatively small, changing the surface temperature by less than 5 K relative to an atmosphere that includes only H$_2$ and H$_2$O.

As shown in Figure~\ref{fig:stellar_spectrum}, the M-dwarf host star (blue line) emits most of its energy in the near-infrared, in stark contrast to the Solar spectrum (red line), which peaks in the visible. Because the incident flux is relatively low in the visible range where Rayleigh scattering cross-sections are large ($\propto \lambda^{-4}$), a purely gaseous atmosphere with Rayleigh scattering alone would have a much lower Bond albedo than under solar irradiation. This implies that additional reflective agents are required to raise the planetary albedo, a point that we further explore in Section~\ref{sec:haze_albedo}.

\begin{figure}
 \centering
 \includegraphics[width=\columnwidth]{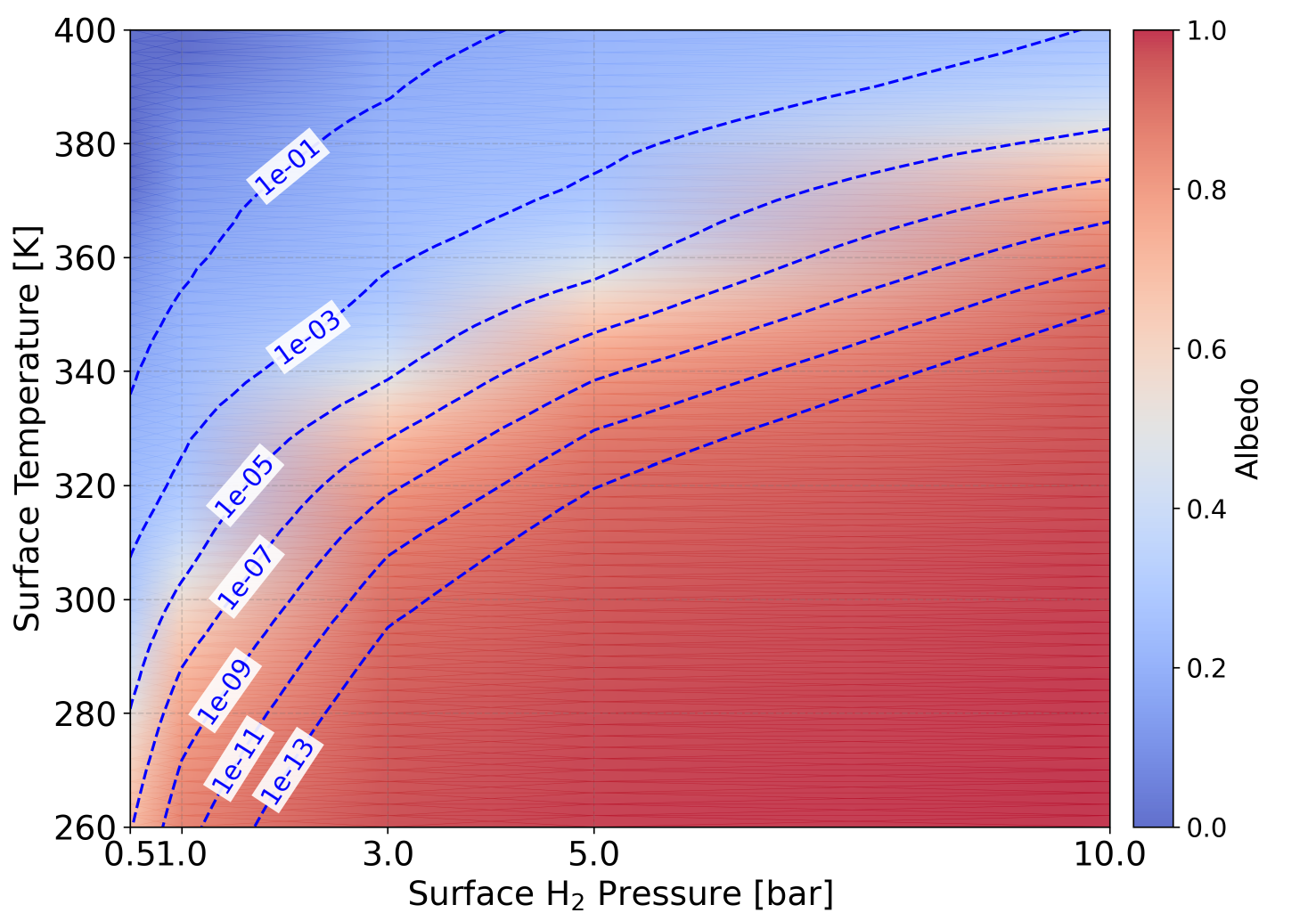}
 \caption{
Surface pressure--temperature diagram with contours of the stratospheric H$_2$O mixing ratio.
The horizontal axis shows the surface pressure, and the vertical axis shows the surface temperature.
Solid curves indicate lines of constant H$_2$O mixing ratio in the stratosphere, evaluated at the cold trap near the tropopause.
 }
 \label{fig:pt_stratospheric_h2o}
\end{figure}

To connect these climatic constraints to the photochemical parameter space, we also examine how the stratospheric H$_2$O mixing ratio varies across our grid of radiative–convective solutions. Figure~\ref{fig:pt_stratospheric_h2o} illustrates representative relationships between surface temperature, H$_2$ surface pressure, and the resulting stratospheric H$_2$O abundance. The stratospheric H$_2$O mixing ratio is diagnosed at the cold trap (tropopause) level from the radiative–convective P–T profiles, assuming that H$_2$O follows saturation up to the cold trap. To facilitate comparison with Figure~\ref{fig:pressure_albedo}, the Bond albedo corresponding to each solution is also indicated by a background colour gradient. Taking the 1 bar cases as an example, as the surface warms and the tropopause temperature increases, the cold trap becomes less efficient, and the stratospheric H$_2$O mixing ratio increases from $\sim 10^{-7}$–$10^{-6}$ in cooler, lower-pressure solutions to $\sim 10^{-2}$–$10^{-1}$ in warmer, higher-pressure regimes approaching the dashed runaway boundary in Figure~\ref{fig:pressure_albedo}.

These trends provide a physical underpinning for the range of stratospheric H$_2$O mixing ratios explored in our photochemical experiments lower-boundary-CO2-free experiments, and clarify which parts of that parameter space are compatible with temperate Hycean climates. A more detailed discussion of how these climatic constraints relate to the CO$_2$ mass balance and ocean buffering is given in Section~\ref{subsec:H2O_CO2_Hycean}.

\section{Discussion} \label{sec:discussion}

\subsection{Impact of Photochemically Induced Vertical Gradients on Spectral Interpretation}
\label{sec:photochemical_saturation}
Our results demonstrate that the vertical distribution of chemical species, particularly the sharp decrease in CH$_4$ due to photodissociation, significantly alters the transmission spectrum structure. As described in Section 3.3, this sharp gradient causes the optical depth to increase abruptly at a specific altitude, leading to a saturation of the transit depth across a broad wavelength range where CH$_4$ absorption dominates. This produces a flat-topped spectral feature (a plateau) that differs from the absorption shape expected from a vertically uniform abundance profile.

This finding has critical implications for atmospheric retrieval analyses. Standard retrieval algorithms often assume vertically uniform mixing ratios (isobaric profiles) to reduce computational complexity. However, applying such uniform models to an atmosphere with strong vertical gradients—typical for Hycean worlds under strong UV irradiation—may lead to biased abundance estimates, so that utilizing photochemically motivated vertical profiles becomes important when interpreting JWST data of sub-Neptunes like K2-18b.

\subsection{Consequences of $R\approx55$ Binning for CO and CO$_2$ Abundance Estimates}
\label{subsec:exoTEDRF_R55}

\begin{figure*}
    \centering
    \includegraphics[width=\textwidth]{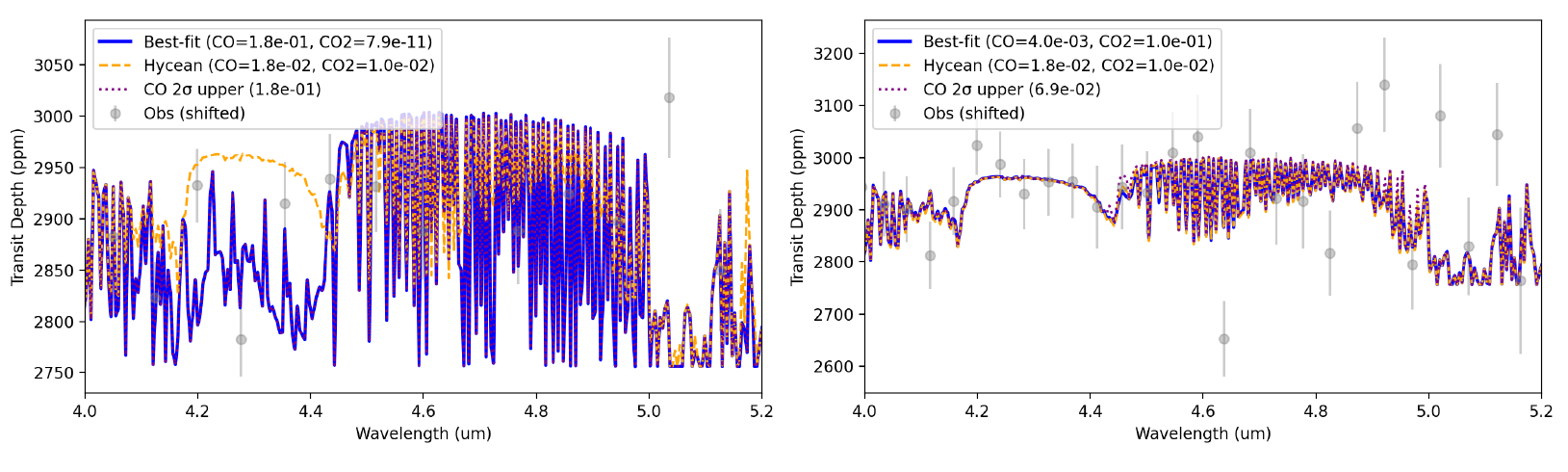}
    \caption{
    Comparison of best-fit Hycean model spectra for the
    \texttt{exoTEDRF} reduction in the CO/CO$_2$-sensitive region.
    Left: R$\approx$55 binning using the wavelength grid
    adopted by \citet{Madhusudhan2023}, applied to the
    \texttt{exoTEDRF} spectra. Right: R$\approx$100 binning constructed
    consistently from the native exoTEDRF pixel-level spectra
    and their associated uncertainties. In both panels the
    solid curves show the best-fit models within our Hycean–
    photochemistry framework, while points with error
    bars show the corresponding binned data. In the
    R$\approx$55 case the fits tend to favour relatively high
    CO and comparatively low CO$_2$ mixing ratios,
    whereas the R$\approx$100 analysis prefers lower CO
    and higher CO$_2$ abundances under otherwise identical
    modelling assumptions.
    }
    \label{fig:exoTEDRF_R55_R100}
\end{figure*}

Figure~\ref{fig:exoTEDRF_R55_R100} compares our Hycean
best-fit spectra for the \texttt{exoTEDRF} reduction in the 4--5.2 ~$\micron$
band when analysed at two different spectral resolutions.
In the left panel, we adopt the R$\approx$55 wavelength grid
used by \citet{Madhusudhan2023}, rebinned from the
\texttt{exoTEDRF} spectra to allow a direct visual comparison
with their analysis. In this configuration, our Hycean
forward models tend to favour solutions with relatively
high CO mixing ratios and comparatively low CO$_2$,
broadly consistent with the CO-rich, CO$_2$-poor
posteriors reported in some low-resolution reanalyses.

In contrast, when we perform a fully self-consistent
analysis at R$\approx$100 using the native \texttt{exoTEDRF}
pixel-level data and error estimates (right panel), the
preferred balance between CO and CO$_2$ changes
qualitatively. Under the same Hycean temperature–
pressure structure and photochemical vertical profiles,
the R$\approx$100 spectra favour solutions with modest
or low CO and enhanced CO$_2$ mixing ratios, in line
with the trends seen for the \texttt{FIREFLy} and \texttt{Eureka!} reductions at the same resolution
(Section~\ref{subsec:CO_results} and
Figure~\ref{fig:pCO_panels}). This behaviour indicates
that the inferred CO/CO$_2$ trade-off is sensitive not
only to the choice of reduction pipeline, but also to how
the native pixel data are binned and weighted.

A plausible explanation is that adopting the
\citet{Madhusudhan2023} R$\approx$55 binning for the
\texttt{exoTEDRF} spectra implicitly imports the weighting
scheme and systematic-error structure appropriate for
the original \texttt{JExoRES} reduction into the \texttt{exoTEDRF}
data. In such a case, the effective uncertainties and
covariances of the R$\approx$55 points may not faithfully
reflect the noise properties of the \texttt{exoTEDRF} pipeline,
so that subtle residual systematics in the 4--5~$\micron$
region can be absorbed preferentially by CO rather than
CO$_2$ in the low-resolution fit. By contrast, our
R$\approx$100 analysis uses a binning and error propagation
procedure tailored to the \texttt{exoTEDRF} pixel data, yielding
a more internally consistent weighting of the CO and
CO$_2$ features.

Crucially, when we restrict attention to the
R$\approx$100 analyses, the inferred preference for
moderately high CO$_2$ mixing ratios becomes qualitatively
similar across all reductions we consider, despite their
independent treatments of background subtraction,
1/$f$ noise, and light-curve systematics. We therefore
regard the R$\approx$100 results as a more robust guide
for interpreting oxidised carbon abundances in Hycean
models than mixed-resolution comparisons based on a
common R$\approx$55 grid. At the same time, the
discrepancy between the R$\approx$55 and R$\approx$100
results for \texttt{exoTEDRF} highlights that current CO
constraints remain sensitive to details of binning and
error modelling, and should not yet be used to rule out
Hycean scenarios. Instead, present CO retrievals are
better viewed as reflecting the limited information
content of the CO band at current signal-to-noise,
and the susceptibility of low-resolution fits to the
adopted binning scheme.

\subsection{Hycean Forward Models Consistent with JWST Transmission Spectra}
\label{subsec:Hycean_bestfit}

\begin{figure*}
    \centering
    \includegraphics[width=\textwidth]{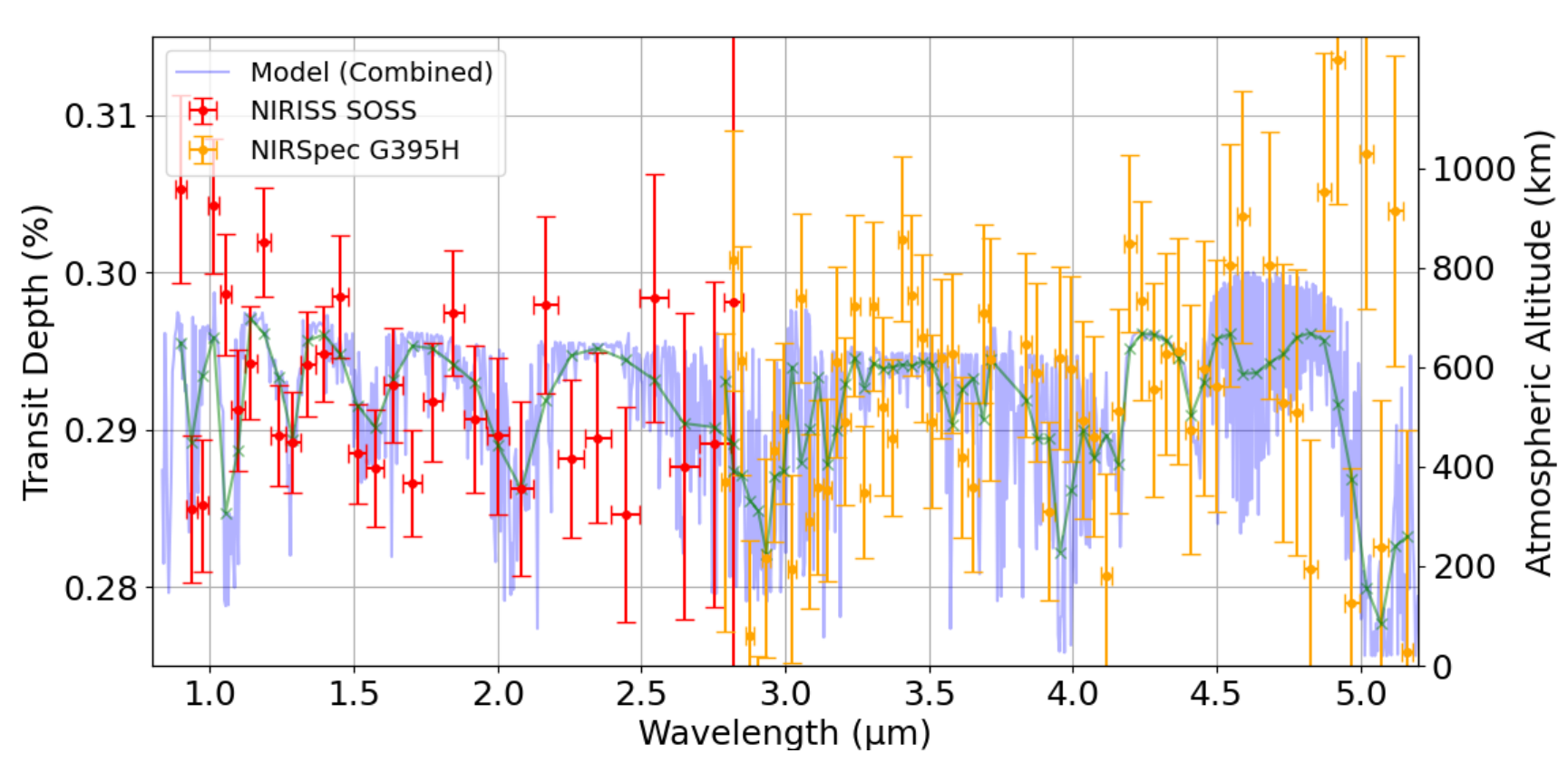}
    \caption{
    Best-fit Hycean transmission spectrum for K2-18b
    compared with the combined JWST NIRISS SOSS and
    NIRSpec G395H data from the \texttt{exoTEDRF} reduction.
    The solid purple line shows the forward model computed
    using the photochemical profiles for
    H$_2$, CH$_4$, CO, CO$_2$, C$_2$H$_2$, and C$_2$H$_6$
    (Figure~\ref{fig:mixing_ratios_co2_case}), together with
    a radiative–convective $P$–$T$ structure consistent with
    a 1~bar H$_2$ envelope over a liquid H$_2$O ocean
    (Section~\ref{sec:climate_stability}). Red and orange
    points with error bars denote the NIRISS SOSS and
    NIRSpec G395H data, respectively, binned at
    R$\approx$25 (NIRISS SOSS) and R$\approx$100 (NIRSpec G395H).
    The green dashed curve shows the same model averaged
    over the observational bins.}
    \label{fig:Hycean_bestfit}
\end{figure*}

Figure~\ref{fig:Hycean_bestfit} presents the full best-fit
Hycean transmission spectrum obtained within our
modelling framework, alongside the combined NIRISS
SOSS and NIRSpec G395H data from the \texttt{exoTEDRF}
reduction. The forward model (solid purple line)
incorporates the altitude-dependent mixing-ratio
profiles of H$_2$, CH$_4$, CO, CO$_2$, C$_2$H$_2$, and
C$_2$H$_6$ predicted by our photochemical calculation for a 1~bar H$_2$
envelope over a liquid H$_2$O ocean
(Figure~\ref{fig:mixing_ratios_co2_case}), together with
a radiative–convective $P$–$T$ profile that avoids a
runaway-greenhouse state for K2-18b
(Section~\ref{sec:climate_stability}). The same model,
averaged over the observational bins, is shown by the
green dashed curve. Red and orange points represent
the NIRISS SOSS and NIRSpec G395H data, respectively,
with NIRISS binned at R$\approx$25 and NIRSpec at
R$\approx$100.

In the NIRISS wavelength range (0.8--2.8~$\micron$),
the Hycean+photochemistry model naturally reproduces
the CH$_4$-dominated band structure, including the
flattened plateaux that arise from photochemical
depletion of CH$_4$ at high altitudes
(Section~\ref{subsubsec:transit_profiles}). This
``photochemical saturation'' behaviour reflects the fact
that, over broad CH$_4$-dominated wavelength intervals,
the effective transit radius is set by nearly the same
altitude just below the CH$_4$ depletion layer. Once the
offset is calibrated in the 2.8--4.0~$\micron$ CH$_4$ band
(Section~\ref{subsubsec:offset_results}), the Hycean
model reproduces both the overall amplitude and the
shape of the near-infrared features.

At longer wavelengths (2.8--5.2~$\micron$), the same
Hycean model remains compatible with the NIRSpec
G395H data when the CO and CO$_2$ abundances are
allowed to vary within the ranges supported by our
photochemical mass-balance and ocean-buffering analysis
(Section~\ref{sec:CO2_H2O_dependence} and
Section~\ref{sec:appendix_CO2mass}). The 4--5~$\micron$
region, which is sensitive to the CO fundamental band
and the strong CO$_2$ band near 4.3~$\micron$, can be
explained by combinations of CO and CO$_2$ that
naturally emerge in our H$_2$–CH$_4$–H$_2$O networks,
without invoking additional species such as DMS. Given
that the DMS absorption bands overlap strongly with
the CO and CO$_2$ features in this region (Figure \ref{fig:opacity_spectra}), and that
current data do not uniquely disentangle these
contributions, we do not find it necessary to include
DMS to reproduce the observed spectrum.

Taken together, the full-spectrum comparison in
Figure~\ref{fig:Hycean_bestfit} shows that a Hycean
atmosphere with an H$_2$ envelope of order 1~bar,
percent-level CH$_4$, and CO$_2$ buffered at
$\sim 10^{-3}$--$10^{-2}$ can reproduce the existing JWST
transmission spectra of K2-18b once photochemical
vertical gradients and a physically motivated offset
calibration are taken into account. While mini-Neptune
interpretations remain viable, our results demonstrate
that Hycean scenarios are likewise consistent with the
data, and that current CO and CO$_2$ constraints are
not yet sufficient to exclude such configurations.

\subsection{Stratospheric $\mathrm{H_2O}$ and Buffered $\mathrm{CO_2}$ in Hycean Interpretations}
\label{subsec:H2O_CO2_Hycean}

As shown in Figure~\ref{fig:pt_stratospheric_h2o}, under Hycean-like conditions with a 1 bar H$_2$ envelope, the stratospheric H$_2$O mixing ratio is generally not extremely low. Across the non–runaway part of the pressure–temperature plane relevant for Hycean scenarios, the H$_2$O abundance at the cold trap is typically $\gtrsim 10^{-7}$, and tends to decrease toward higher surface pressures and lower surface temperatures. For warm solutions with Bond albedos $\lesssim 0.5$, photolysis of H$_2$O supplies abundant OH radicals, which in turn enhances CO$_2$ production via the CO + OH pathway given by Equation~(\ref{eq:CO_OH_CO2}).

The left panel of Figure~\ref{fig:co2_h2o_massbalance} shows, for a given stratospheric H$_2$O abundance, the atmospheric CO$_2$ mixing ratio that can be maintained over gigayear timescales when ocean uptake is taken into account. Comparing this with Figure~\ref{fig:pt_stratospheric_h2o}, we see that for Hycean-like stratospheric H$_2$O levels ($\gtrsim 10^{-5}$), the CO$_2$ production flux is large enough that keeping the atmospheric CO$_2$ mixing ratio at $\sim 10^{-3}$---i.e., at a level where CO$_2$ features remain muted and comparable to the mini-Neptune interpretation proposed by \citet{Schmidt2025}---requires oceans that are substantially alkaline. In our ocean model, achieving CO$_2 \sim 10^{-3}$ under such Hycean conditions is only possible if the global ocean is clearly more alkaline than the present-day Earth ocean (pH $\sim 8$), corresponding to pH values firmly in the mildly to strongly alkaline regime. This suggests that reproducing the mini-Neptune-like CO$_2 \sim 10^{-3}$ level within a Hycean framework is, in principle, possible, but demands ocean chemistries that are considerably more alkaline than can be easily justified by direct Earth analogy.

In these experiments, the stratospheric temperature is treated as equal to the skin temperature inferred solely from the planet’s outgoing longwave radiation, without explicitly accounting for radiative cooling or heating by additional atmospheric constituents. Therefore, the actual water abundance at the tropopause could be lower if the tropopause temperature is reduced by these effects. If the stratosphere can be kept significantly drier than in the standard Hycean solutions considered here---with H$_2$O mixing ratios $\lesssim 10^{-6}$–$10^{-7}$---the situation changes. In this regime, as illustrated in the left panel of Figure~\ref{fig:co2_h2o_massbalance}, the photochemical CO$_2$ production flux is strongly suppressed, and the range of ocean conditions that can maintain CO$_2$ mixing ratios $\lesssim 10^{-3}$ over gigayear timescales becomes much broader, including more weakly alkaline oceans with pH values closer to that of present-day Earth.

Moreover, if one places some weight on the CO$_2$-rich
solutions preferred by several reductions in
Figure~\ref{fig:pCO_panels} and
Table~\ref{tab:CO_stats}, then even more nearly neutral
ocean chemistries (with pH only slightly above or
comparable to modern Earth) can also be accommodated
within the same mass-balance framework.

A key consideration in this discussion is that surface warming affects climate and ocean chemistry in opposite directions from the standpoint of the CO$_2$ mass balance. As the surface warms, the ocean depth and carbon storage capacity tend to increase, which acts to lower atmospheric CO$_2$. At the same time, however, the tropopause temperature rises and the stratospheric H$_2$O mixing ratio increases, thereby enhancing the photochemical CO$_2$ production flux. In this work, we have treated these feedbacks in a simplified manner. A fully self-consistent analysis that simultaneously couples climate, ocean depth, and CO$_2$ production would likely further narrow the range of Hycean conditions that are both CO$_2$-poor and compatible with the current observational constraints.

\subsection{Mass Balance of $\mathrm{CH_4}$ and $\mathrm{H_2}$ in a Hycean Atmosphere}\label{sec:CH4,H2}

As described in Section~\ref{sec:photochemi}, for a 1~bar H$_2$ envelope containing CH$_4$ with a mixing ratio of 1\%, the characteristic photochemical lifetime of CH$_4$ is estimated to be $\sim 1.2 \times 10^{7}$~yr.

In contrast, the dominant sink of H$_2$ is atmospheric escape driven by high-energy stellar irradiation. Using the CH$_4$-cooling-limited hydrodynamic hydrogen escape formulation of \citet{Yoshida2020}, the mass-loss rate of H$_2$ from a 1~bar H$_2$ envelope on K2-18b is estimated to be $\sim 2.64 \times 10^{4}$~kg~s$^{-1}$. On the other hand, the net H$_2$ production rate obtained from our photochemical calculations in Section~\ref{sec:photochemi} is $\sim 2.65 \times 10^{3}$~kg~s$^{-1}$. The resulting net H$_2$ loss rate is therefore $\sim 2.38 \times 10^{4}$~kg~s$^{-1}$. This net photochemical H$_2$ production arises from the consumption of CH$_4$ and H$_2$O along the reaction pathways highlighted in Figure~\ref{fig:reaction_network}. Comparing the net loss rate with the total H$_2$ mass of the 1~bar envelope, $2.80 \times 10^{19}$~kg, yields a characteristic depletion timescale for the H$_2$ envelope of $\sim 4 \times 10^{7}$~yr.

Such a short timescale implies that, if K2-18b currently hosts a Hycean-type atmosphere with a $\sim 1$~bar H$_2$ envelope and a CH$_4$ mixing ratio of 1\%, then either (i) the planet is being observed in a transient evolutionary phase, or (ii) there exist sufficiently strong internal sources of both CH$_4$ and H$_2$ that can compensate for atmospheric escape.

One plausible mechanism capable of continuously replenishing both CH$_4$ and H$_2$ is the thermal decomposition of primordial organic material sequestered in the planetary interior. Evolution models suggest that sub-Neptunes like K2-18b may have formed beyond the snow line, accreting substantial amounts of water ice and comet-like carbonaceous material \citep{Rogers2011,Lopez2014}. For example, the bulk composition of solid dust in Halley's comet is rich in hydrogen and carbon, with mass fractions of H $\sim 7.8$\% and C $\sim 37.5$\% \citep{Jessberger1988}. In such a scenario, a large reservoir of organics could be trapped within a high-pressure ice mantle or near the rocky core--ice boundary, consistent with Hycean interior-structure models \citep{Madhusudhan2021}. When interior temperatures become sufficiently high, this material would undergo pyrolysis and release volatile species such as CH$_4$ and H$_2$. Studies of volatile dissolution and outgassing under high-temperature, high-pressure conditions (e.g., \citealt{Kite2020}) further suggest that continuous atmospheric supply could be maintained in a manner analogous to magma--atmosphere systems, even if the volatile reservoir resides beneath a thick ice layer.

We now make an order-of-magnitude estimate of the required scale of such interior sources. The key point is that, if the internal volatile flux is large enough to offset the net H$_2$ loss of $\sim 2.38 \times 10^{4}$~kg~s$^{-1}$, then the associated carbon flux from the same organic reservoir would comfortably exceed the rate needed to replenish CH$_4$ against photochemical destruction. Typical comet-like organics are rich in both H and C \citep{Jessberger1988}, so any internal reservoir capable of supplying enough hydrogen (in the form of H$_2$ or its precursors) to balance escape would, by construction, also release sufficient carbon to sustain the observed CH$_4$.

Assuming an average hydrogen mass fraction of 7.8\% in the organic material, compensating a net H$_2$ loss of $\sim 2.38 \times 10^{4}$~kg~s$^{-1}$ requires an organic processing rate of $\sim 3.04 \times 10^{5}$~kg~s$^{-1}$. If this supply were maintained over 3~Gyr, the total processed organic mass would be $\sim 2.87 \times 10^{22}$~kg, corresponding to $\sim 560$~ppm of K2-18b's total mass (8.6~M$_\oplus$). Although this is larger than Earth's bulk carbon inventory (tens to hundreds of ppm; \citealt{Dasgupta2010}), it remains plausible in a scenario where a sub-Neptune forms beyond the snow line and accretes large amounts of volatile-rich bodies \citep{Rogers2011,Lopez2014}. In such a case, the carbon released alongside hydrogen would exceed the CH$_4$ photochemical loss rate inferred in Section~\ref{sec:photochemi}, allowing atmospheric CH$_4$ to be replenished over geological timescales.

An additional and potentially important internal source of H$_2$ is the reaction of water with reduced iron in the planetary interior. K2-18b likely possesses a rocky core beneath a high-pressure ice mantle \citep{Madhusudhan2021}, and water--rock interactions at this boundary can oxidise ferrous iron in silicate rocks while producing H$_2$ \citep{Kite2020}. This process can be further enhanced if water-rich comets and planetesimals, which deliver both H$_2$O and Fe, have been accreted in large quantities. Thus, the H$_2$ replenishment needed to counteract atmospheric escape may arise from a combination of thermal decomposition of accreted organics and abiotic H$_2$ production at the ice--mantle boundary. If these interior sources are sufficiently strong, they can, in principle, sustain both the thin H$_2$ envelope and the observed CH$_4$ abundance against photochemical destruction and atmospheric escape over geological timescales.

\subsection{Haze Production and the Feasibility of High-Albedo Hycean Atmospheres}
\label{sec:haze_albedo}

Our radiative–convective calculations (Section~\ref{sec:climate_stability}; Figure~\ref{fig:pressure_albedo}) indicate that, under K2-18b's incident flux, a H$_2$-rich atmosphere requires a planetary Bond albedo of order $A_{\mathrm{B}}\sim 0.3$ to maintain surface temperatures of $T_{\mathrm{s}}\simeq 320$--$330$~K and avoid a runaway-greenhouse regime. This raises the question of whether photochemically produced hazes can contribute significantly to such an albedo.

Using the C$_3$H$_8$ loss rates from our photochemical simulations, we estimate a total C$_3$H$_8$ loss flux of $F_{\mathrm{C_3H_8,loss}}\simeq 1.8\times 10^{4}$~kg~s$^{-1}$. If we assume that only about 1\% of this loss ultimately proceeds toward polymerisation and condensation into haze particles, the resulting organic haze precursor flux is $F_{\mathrm{haze}}\sim 10^{2}$~kg~s$^{-1}$. For K2-18b’s surface area, this corresponds to a globally averaged column production rate of $f_{\mathrm{haze}}\sim 10^{-14}$–$10^{-13}$~kg~m$^{-2}$~s$^{-1}$, which can sustain haze column masses $m_{\mathrm{col}}\sim 10^{-5}$–$10^{-4}$~kg~m$^{-2}$ for residence times of 10–100~yr.

Combined with mass extinction coefficients $\kappa_\lambda\sim 10^{3}$–$10^{4}$~m$^{2}$~kg$^{-1}$ for Titan-like organic aerosols, these column masses yield visible optical depths $\tau_\lambda\sim 0.01$–1. Order-of-magnitude comparisons with Titan suggest that such optically thin to moderately thick haze layers can plausibly support planetary albedos of order $A_{\mathrm{B}}\sim 0.2$–0.3, especially when supplemented by underlying condensate clouds. In Section~\ref{sec:appendix_haze}, we provide the detailed derivation of these estimates based on the full C$_3$H$_8$ loss budget and residence times.

We also find that, in additional experiments with higher CH$_4$ mixing ratios (15\%, consistent with the observational constraint CH$_4<30$\% from \citealt{Schmidt2025}), the fraction of C$_3$H$_8$ loss proceeding through polymerising channels increases with altitude and can reach several per cent to $\sim 10$--20\% of the total C$_3$H$_8$ loss in the upper atmosphere. This trend indicates that in more CH$_4$-rich Hycean atmospheres the effective haze production flux may be larger than the conservative 1\% assumption adopted above, making photochemical hazes an even more efficient contributor to the planetary albedo.


\section{Conclusion}
\label{sec:conclusion}

In this work, we have assessed the self-consistency of
Hycean interpretations for K2-18b by combining a
one-dimensional photochemical model, radiative--convective
equilibrium calculations, and forward modelling of
transmission spectra. In particular, we used a Hycean +
photochemistry framework to calibrate the offset in the
CH$_4$-dominated 2.8--4.0\,$\micron$ band and, under this
condition, explored the allowed ranges of CO and CO$_2$
in the 4--5\,$\micron$ band, thereby evaluating to what
extent Hycean models can reproduce the current JWST
spectra.

Our photochemical calculations show that in an
H$_2$--CH$_4$--H$_2$O atmosphere CO is naturally driven
toward mixing ratios of order 1--2\% for both CO2-bearing and CO2-free lower-boundary conditions and over a range of stratospheric H2O abundances. Such CO levels are difficult to
suppress with simple ocean dissolution or near-surface
sinks and appear formally in tension with some
CO-poor retrieval results. However, the probability
distributions of CO inferred from our CH$_4$-anchored
offset analysis across multiple reductions allow
comparatively high CO abundances, and the limited
information content of the 4--5\,$\micron$ band, together
with the sensitivity of CO constraints to binning and
reduction choices, means that current non-detections of
CO cannot be used to rule out Hycean scenarios.

For CO$_2$, we find that even without imposing a
lower-boundary CO$_2$ flux, H$_2$--CH$_4$--H$_2$O
photochemistry produces a finite CO$_2$ flux that, when
combined with an ocean-buffering model, can be
balanced at atmospheric mixing ratios of
$\sim 10^{-3}$--$10^{-2}$. This range is consistent with
the upper limits from previous JWST retrievals and fits
comfortably within the CO$_2$ abundances obtained by
our Hycean-based spectral analysis. Thus, Hycean
atmospheres can remain compatible with the apparent
non-detection of CO$_2$.

From the standpoint of the mass balance of CH$_4$ and
H$_2$, our results indicate that a $\sim$1\,bar H$_2$
envelope with percent-level CH$_4$ requires a balance
between interior gas supply and atmospheric escape on
gigayear timescales. This suggests that a Hycean
atmosphere on K2-18b, if present, is unlikely to be a
static equilibrium state, but rather a dynamic steady
state in which interior outgassing, photochemistry, and
escape are approximately balanced.

In comparing model and data, we find that Hycean +
photochemistry models naturally reproduce the
CH$_4$-dominated structure of the NIRISS SOSS spectrum
over 0.8--2.8\,$\micron$, including the plateau-like
features caused by photochemical depletion of CH$_4$
(photochemical saturation). When CO and CO$_2$ are
varied within the ranges permitted by photochemistry
and ocean buffering, the NIRSpec G395H spectrum in the
4--5\,$\micron$ band can also be explained without
invoking additional species such as DMS. 

Taken together, these results show that a Hycean
atmosphere with a $\sim$1\,bar H$_2$ envelope,
percent-level CH$_4$, and CO$_2$ buffered at
$\sim 10^{-3}$--$10^{-2}$ can reproduce the current JWST
transmission spectra once photochemical vertical
profiles and physically motivated offsets are taken into
account. At the same time, mini-Neptune interpretations
remain consistent with the data, and present constraints
on CO and CO$_2$ alone are insufficient to decisively
favour or exclude either scenario. We therefore conclude
that the possibility that K2-18b hosts a Hycean
atmosphere remains open at the current stage.

\appendix

\section{Order-of-Magnitude Estimate of Long-Term $\mathrm{CO_2}$ Mass Balance}
\label{sec:appendix_CO2mass}

In this Appendix, we summarise a simple order-of-magnitude estimate for the long-term mass balance of CO$_2$ in a Hycean atmosphere–ocean system. The goal is to assess under what conditions a CO$_2$-poor atmosphere can be maintained over gigayear timescales when photochemistry continuously produces CO$_2$ from reduced carbon species.

\subsection{Photochemical $\mathrm{CO_2}$ Production and Atmospheric Source Term}

From the photochemical experiments lower-boundary-CO2-free experiments (Section~\ref{sec: PM}), we obtain a net CO$_2$ production rate
\begin{equation}
F_{\mathrm{CO_2}} \quad [\mathrm{kg\,s^{-1}}],
\end{equation}%
representing the integrated mass flux of CO$_2$ produced throughout the atmosphere by reactions such as
\begin{equation}
\mathrm{CO} + \mathrm{OH} \rightarrow \mathrm{CO_2} + \mathrm{H}.
\end{equation}%
Over a timescale $t$ comparable to the age of the system (e.g., $t \sim 3\times 10^{9}\,\mathrm{yr}$), the total mass of CO$_2$ produced photochemically is
\begin{equation}
M_{\mathrm{CO_2,prod}} = F_{\mathrm{CO_2}}\, t.
\end{equation}%
For the values of $F_{\mathrm{CO_2}}$ obtained in our CO$_2$-free runs, $M_{\mathrm{CO_2,prod}}$ is of order $10^{20}$–$10^{21}\,\mathrm{kg}$ over $t \sim 3$~Gyr.

\subsection{Oceanic Dissolution and Carbonate Buffering}

We assume that the planet hosts a global liquid-water ocean of volume
\begin{equation}
V_{\mathrm{ocean}} = f\, V_{0},
\end{equation}%
where $V_{0}$ is a reference ocean volume corresponding to a fiducial Hycean configuration, and $f$ is an ocean-depth parameter that scales the total liquid volume. For K2-18b-like parameters, $V_{0} \sim 3\times10^{23}\,\mathrm{L}$ is a representative value. Internal structure models for water-rich sub-Neptunes similar to K2-18b \citep[e.g.,][]{Madhusudhan2021} further suggest that, for a given bulk composition and gravity, variations of the surface temperature within the temperate range relevant to this work (typically $T_0 \sim 300$--350~K) change the depth of the outer liquid H$_2$O layer by factors of at most a few rather than by orders of magnitude. This behaviour is illustrated, for example, by the ocean-depth calculations shown in their Figure 10, where modest changes in $T_0$ around 300~K shift the ocean depth substantially, but still within the same order of magnitude for fixed mass and composition. Motivated by these results, in this study we focus on cases in which the ocean volume differs from the reference Hycean configuration by at most a factor of a few, i.e.\ we treat $f$ values within a range of order several around unity as the physically plausible regime.

At the surface temperature $T_{\mathrm{s}}$, the concentration of dissolved molecular CO$_2$(aq) in the ocean is related to the atmospheric CO$_2$ partial pressure $P_{\mathrm{CO_2}}$ by Henry’s law,
\begin{equation}
[\mathrm{CO_2(aq)}] = k_{\mathrm{H}}(T_{\mathrm{s}})\, P_{\mathrm{CO_2}},
\end{equation}%
where $k_{\mathrm{H}}(T_{\mathrm{s}})$ is the Henry constant (mol\,L$^{-1}$\,Pa$^{-1}$). In an aqueous solution, CO$_2$(aq) participates in the carbonate equilibria
\begin{equation}
\mathrm{CO_2(aq)} + \mathrm{H_2O} \rightleftharpoons \mathrm{H^+} + \mathrm{HCO_3^-},
\end{equation}
\begin{equation}
\mathrm{HCO_3^-} \rightleftharpoons \mathrm{H^+} + \mathrm{CO_3^{2-}},
\end{equation}%
so that the total dissolved inorganic carbon (DIC) is
\begin{equation}
\mathrm{DIC} = [\mathrm{CO_2(aq)}] + [\mathrm{HCO_3^-}] + [\mathrm{CO_3^{2-}}].
\end{equation}%
We define a carbonate enhancement factor $\beta$ as
\begin{equation}
\beta(T_{\mathrm{s}}, \mathrm{pH}) \equiv 
1 + \frac{[\mathrm{HCO_3^-}]}{[\mathrm{CO_2(aq)}]} + \frac{[\mathrm{CO_3^{2-}}]}{[\mathrm{CO_2(aq)}]},
\end{equation}%
so that
\begin{equation}
\mathrm{DIC} = \beta\, [\mathrm{CO_2(aq)}].
\end{equation}%
For mildly alkaline solutions (Earth-like pH $\sim 8$), $\beta \sim 10^{2}$, while more alkaline conditions can yield $\beta \sim 10^{2}$–$10^{3}$.

The total mass of dissolved molecular CO$_2$(aq) in the ocean is
\begin{equation}
M_{\mathrm{CO_2(aq)}} = [\mathrm{CO_2(aq)}]\, M_{\mathrm{CO_2}}\, V_{\mathrm{ocean}},
\end{equation}%
where $M_{\mathrm{CO_2}} = 44\times 10^{-3}\,\mathrm{kg\,mol^{-1}}$ is the molar mass of CO$_2$ and we adopt a liquid density of $\sim 1\,\mathrm{kg\,L^{-1}}$. The total mass of dissolved inorganic carbon is then
\begin{equation}
M_{\mathrm{C,tot}} = \beta\, M_{\mathrm{CO_2(aq)}}.
\end{equation}

Using Henry’s law to express $[\mathrm{CO_2(aq)}]$ in terms of $P_{\mathrm{CO_2}}$, we obtain
\begin{equation}
M_{\mathrm{C,tot}} = \beta\, k_{\mathrm{H}}(T_{\mathrm{s}})\, P_{\mathrm{CO_2}}\, M_{\mathrm{CO_2}}\, V_{\mathrm{ocean}}.
\end{equation}%
For convenience, we write
\begin{equation}
M_{\mathrm{C,tot}} \simeq 
\left( \beta\, f \right)\, C(T_{\mathrm{s}})\, P_{\mathrm{CO_2}},
\end{equation}%
where $C(T_{\mathrm{s}})$ collects the temperature-dependent factors $k_{\mathrm{H}}(T_{\mathrm{s}})$, $M_{\mathrm{CO_2}}$, and $V_{0}$.

\section{Mass-Balance Condition and Allowed $\mathrm{CO_2}$ Abundances}

Long-term mass balance requires that the total amount of oxidised carbon stored in the ocean–interior system be at least comparable to the integrated photochemical production over the timescale $t$,
\begin{equation}
M_{\mathrm{C,tot}} \gtrsim M_{\mathrm{CO_2,prod}} = F_{\mathrm{CO_2}}\, t.
\end{equation}%
Using the expression for $M_{\mathrm{C,tot}}$ above, this condition can be written as a lower bound on $\beta f$ for a given atmospheric CO$_2$ partial pressure $P_{\mathrm{CO_2}}$ and photochemical source $F_{\mathrm{CO_2}}$,
\begin{equation}
\beta f \gtrsim \frac{F_{\mathrm{CO_2}}\, t}{C(T_{\mathrm{s}})\, P_{\mathrm{CO_2}}}.
\end{equation}

Equivalently, for fixed $(\beta, f)$ and $F_{\mathrm{CO_2}}$, we can solve for the atmospheric CO$_2$ partial pressure (or mixing ratio) that is compatible with long-term mass balance,
\begin{equation}
P_{\mathrm{CO_2}} \lesssim 
\frac{\beta f\, C(T_{\mathrm{s}})}{F_{\mathrm{CO_2}}\, t}.
\end{equation}%
Adopting representative values for K2-18b-like conditions (e.g., $t \sim 3\times10^9$~yr, $V_0 \sim 3\times10^{23}$~L, and $k_H(T_s)$ appropriate for $T_s \sim 300$--320~K), and using the CO$_2$ production fluxes from our photochemical experiments lower-boundary-CO2-free experiments, we find that mildly to strongly alkaline oceans with $\beta \sim 10^2$--$10^3$ and ocean depths within a factor of a few of the reference case ($f \sim 1$) can buffer an oxidised carbon inventory comparable to $M_{\rm CO_2,prod}$ while maintaining atmospheric CO$_2$ mixing ratios at the $\sim 10^{-3}$ level.

We emphasise that this estimate is intentionally simplified and neglects additional sinks in the high-pressure ice mantle or rocky core. Nevertheless, it captures the key scaling: for a given photochemical source $F_{\mathrm{CO_2}}$ and timescale $t$, maintaining a CO$_2$-poor atmosphere requires either a sufficiently large product $\beta f$ (efficient oceanic buffering in a deep and/or strongly alkaline ocean) or additional long-term sinks in the planet’s interior that can store a comparable mass of oxidised carbon.

\section{Order-of-Magnitude Estimate of Organic Haze Production}
\label{sec:appendix_haze}

In this Appendix we provide the detailed derivation of the organic haze production estimates used in Section~\ref{sec:haze_albedo}. We base our analysis on the C$_3$H$_8$ loss budget from our photochemical simulations for the fiducial Hycean setup.

\subsection{$\mathrm{C_3H_8}$ Loss Flux and Polymerising Branching Ratio}

In the fiducial model, the total C$_3$H$_8$ loss flux integrated over the atmosphere is
\begin{equation}
 F_{\mathrm{C_3H_8,loss}} \simeq 1.77\times 10^{4}\ \mathrm{kg\ s^{-1}}.
\end{equation}%
Following the approach of \citet{Yoshida2024}, we classify the C$_3$H$_8$ loss channels into three categories: (i) oxidation and H-abstraction reactions (primarily C$_3$H$_8$ + OH $\rightarrow$ C$_3$H$_7$ + H$_2$O), (ii) photolysis reactions that fragment C$_3$H$_8$ into lighter hydrocarbons, and (iii) polymerising reactions that extend the C$_2$–C$_3$ hydrocarbon network towards higher carbon numbers.

A representative polymerising reaction in our network is
\begin{equation}
 \mathrm{C_2H} + \mathrm{C_3H_8} \rightarrow \mathrm{C_2H_2} + \mathrm{C_3H_7},
\end{equation}%
which couples C$_2$H radicals to C$_3$H$_8$ and injects carbon into higher-order hydrocarbon pathways. Analysis of the our photochemical diagnostic output for C$_3$H$_8$ loss shows that, in the lower and middle atmosphere (0--600~km) of the 1\% CH$_4$ case, this polymerising channel typically accounts for of order 1\% of the total C$_3$H$_8$ loss. We therefore adopt a representative polymerising branching ratio
\begin{equation}
 f_{\mathrm{pol}}^{(\mathrm{C_3H_8})} \sim 10^{-2},
\end{equation}%
consistent with the order-of-magnitude estimate in \citet{Yoshida2024}. Assuming that this fraction of the C$_3$H$_8$ loss ultimately feeds the growth of higher hydrocarbons and solid organic particles, the effective haze precursor flux is
\begin{equation}
 F_{\mathrm{haze}} = f_{\mathrm{pol}}^{(\mathrm{C_3H_8})} F_{\mathrm{C_3H_8,loss}} \simeq 1.8\times 10^{2}\ \mathrm{kg\ s^{-1}}.
\end{equation}

For a planetary radius $R_{\mathrm{p}}\simeq 2.6\,R_\oplus$, the surface area is
\begin{equation}
 4\pi R_{\mathrm{p}}^{2} \simeq 3.4\times 10^{15}\ \mathrm{m^{2}},
\end{equation}
so that the globally averaged column production rate is
\begin{equation}
 f_{\mathrm{haze}} = \frac{F_{\mathrm{haze}}}{4\pi R_{\mathrm{p}}^{2}} \sim 5\times 10^{-14}\ \mathrm{kg\ m^{-2}\ s^{-1}}.
\end{equation}

\subsection{Haze Column Mass and Optical Depth}

Assuming an effective residence time $\tau_{\mathrm{res}}$ of haze particles in the atmosphere, the steady-state haze column mass $m_{\mathrm{col}}$ can be approximated as
\begin{equation}
 m_{\mathrm{col}} \simeq f_{\mathrm{haze}}\,\tau_{\mathrm{res}}.
\end{equation}%
Microphysical models and observations of Titan’s haze \citep[e.g.,][]{Trainer2006,Pavlov2001,Pavlov2003} suggest residence times of $\tau_{\mathrm{res}}\sim 10$--100~yr in the stratosphere, i.e. $\sim 3\times 10^{8}$--$3\times 10^{9}$~s. Adopting similar values for Hycean conditions yields
\begin{align}
 m_{\mathrm{col}}(10\ \mathrm{yr}) &\sim 1.6\times 10^{-5}\ \mathrm{kg\ m^{-2}},\\
 m_{\mathrm{col}}(100\ \mathrm{yr}) &\sim 1.6\times 10^{-4}\ \mathrm{kg\ m^{-2}}.
\end{align}

The corresponding visible optical depths can be estimated using mass extinction coefficients $\kappa_\lambda$ for Titan-like organic aerosols. Laboratory measurements of Titan tholins \citep{Khare1984} and radiative transfer modelling \citep[e.g.,][]{Trainer2006,Pavlov2001,Pavlov2003} indicate $\kappa_\lambda\sim 10^{3}$--$10^{4}$~m$^{2}$~kg$^{-1}$ in the visible. Thus,
\begin{equation}
 \tau_\lambda = \kappa_\lambda m_{\mathrm{col}} \sim 0.01\text{--}1,
\end{equation}%
i.e. from optically thin to moderately thick haze layers, consistent with the range required to have a substantial impact on planetary albedo.

Finally, we note that in additional experiments with higher CH$_4$ mixing ratios (up to 15\%, still consistent with the observational constraint CH$_4<30$\% from \citealt{Schmidt2025}), the fraction of C$_3$H$_8$ loss proceeding through polymerising channels increases with altitude and can reach several per cent to $\sim 10$--20\% of the total C$_3$H$_8$ loss in the upper atmosphere. This trend suggests that in more CH$_4$-rich Hycean atmospheres the effective haze production flux may be larger than the conservative 1\% assumption adopted here, implying that the optical depths and associated albedo contributions derived above should be regarded as lower bounds.

\begin{acknowledgments}
T.F. performed the majority of the calculations presented in this study. M.S. and T.Y. contributed to the configuration and setup of the PROTEUS model. K.K. provided overall supervision and theoretical guidance. 
\end{acknowledgments}

\section*{DATA AVAILABILITY}

The complete chemical reaction network used in this study, including all rate coefficients, is available as a machine-readable file on Zenodo \citep{OurZenodo2026}.
Other data products that support the findings of this work are available from the corresponding author upon reasonable request.


\bibliographystyle{aasjournal} 
\bibliography{references}    

\end{document}